\documentclass[12pt]{article}
\usepackage[latin1]{inputenc}
\usepackage{amsmath}
\usepackage{amsfonts}
\usepackage[usenames,dvipsnames]{color}
\usepackage{graphicx,psfrag,booktabs}
\usepackage{lineno}
\usepackage{appendix}
\begin{document}
\setlength{\baselineskip}{0.14in}

\title{Depletion of moderately volatile elements by pebble accretion in Earth-like planets }
\author{Peter L. Olson, Zachary D. Sharp, Susmita Garai\\
	    Earth and Planetary Sciences, University of New Mexico} 
\maketitle

\begin{abstract}
Protoplanets growing by pebble accretion capture massive hydrogen-helium atmospheres from the surrounding nebula. Pebbles settling through such atmospheres continuously release gravitational potential energy, heating both the atmosphere and the pebbles. Under these conditions, atmosphere temperatures above large protoplanets are sufficiently high to melt silicate pebbles, support long-lived magma oceans, and drive evaporation of volatile species. Because these atmospheres are open to the nebula, some amount of volatile loss is inevitable.
Here we analyze the depletion of moderately volatile elements from terrestrial protoplanets undergoing pebble accretion. We consider chondrule-size silicate pebbles enriched in Si, Na, K, and Zn relative to Earth, settling through a hydrogen-helium-rich atmosphere containing these same volatiles.  We show that volatile depletion depends critically on protoplanet mass, the timescale of atmosphere exhaust, and the pebble composition. The protoplanetary mass effect is especially strong. For exhaust timescales of a few centuries, we find that substantial depletion of Zn begins around 0.4 Earth mass, and for Na and K around 0.6 Earth mass, with negligible depletion of these elements at smaller masses.  Using a pebble composition that matches Earth's major element abundances, broad agreement with Earth's depletion trend for moderately volatile elements is found by merging a large (approximately 0.7 Earth mass) volatile-depleted target protoplanet with one or more smaller, less-depleted impactors.   
\end{abstract}


$*$ Corresponding author: Peter L. Olson {\it peterleeolson@unm.edu; Earth and Planetary Sciences, Northrop Hall, University of New Mexico, Albuquerque NM 87131}

\section{Introduction} 
Planetary bodies embedded within a protoplanetary disk naturally acquire atmospheres consisting of nebular gas in amounts proportional to their mass. These atmospheres are mostly hydrogen and helium, but may also contain volatiles derived from sublimation of nebular dust (Browers et al., 2018), evaporation from accreting pebbles (Wang et al., 2023), as well as evaporation from the protoplanet surface (Steinmeyer and Johansen, 2024). Because such atmospheres are open systems (Ormel et al., 2015; Bethune and Rafikov, 2019a; Moldenhauer et al., 2022), volatiles produced these ways can be recycled back into the disk and lost from the protoplanet. 

Terrestrial protoplanets growing by pebble accretion are prone to this type of volatile loss. The pebble accretion mechanism relies on nebular gas to provide enough drag so that small solids (the pebbles) can be efficiently captured (Ormel and Klahr, 2010; Morbidelli and Nesvorny, 2012). With an effectively infinite reservoir of nebular gas to draw upon, embedded terrestrial protoplanets can maintain voluminous high-temperature atmospheres, where significant amounts of sublimation, evaporation, and volatile exhaust can occur (Johansen et al., 2023a,b).  Furthermore, pebble accretion is capable of building large terrestrial protoplanets in a few million years (Levison et al., 2015; Johansen et al., 2021; Chambers, 2023), within the nominal lifetime of protoplanetary disks. There is cosmochemical evidence that proto-Earth may well have been largely built within this timeframe (Schiller et al., 2020; Onyett et al., 2023). 

Volatile loss during pebble accretion is one of several early solar system processes that can affect the final volatile composition of a planet.  
Thermal processing mechanisms prior to planet formation includes sublimation in the high-temperature portion of the protoplanetary disk (Li et al., 2021), transient heating by shock waves (Colmenares et al., 2024), and hydrodynamic escape following collisions of small planetesimals (Hin et al., 2017; Braukm\"{u}ller et al., 2019). In addition, later-stage mass addition via impacts may add or remove volatiles, depending on the impactor compositions and the physical conditions at the time of collision (Sharp and Olson, 2022; Grewal et al., 2024; Lock and Stewart, 2024).  

Here we focus on the volatile evolution of a terrestrial protoplanet that grows initially by pebble accretion, then later adds volatiles contained in one or more large impactors. Our model includes atmosphere evaporation kinetics for the accreting pebbles, the protoplanet, and its impactors, along with atmosphere exhaust.  We derive curves of relative abundance versus protoplanet mass for weakly volatile Si and the moderately volatile elements Na, K, and Zn. Comparisons with Earth's abundances of these elements places constraints on the composition and mass of pebbles needed to build the proto-Earth and its impactors this way, and on the exhaust timescale of proto-Earth's nebular atmosphere. 

The general approach adopted in this paper has points in common with a recent study by Wang et al. (2025), who model moderately volatile element depletion trends in Earth and Mars, assuming the two planets were built by different combinations of pebble accretion, planetesimal accretion, and giant impacts. In a later section we summarize the similarities and differences between the Wang et al. (2025) approach and ours, and highlight some important points of agreement.  

\section{Volatile abundances and depletion trends}
Table 1 gives abundances of Mg, Si, Na, K, and Zn for the bulk Earth from Wang et al. (2018), CI chondrites from Lodders (2021), bulk Mars from Yoshizaki and McDonough (2020), and a pebble model from Garai et al. (2025a). 
The pebble model consists of a mixture of chondrules, CAIs (calcium-aluminum inclusions), AOAs (Amoeboid Olivine Aggregates), and metal grains, in proportions that match Earth's major element abundances to within 5-7\% rms. The proportions of each component are constrained so that 60\% of the pebbles by mass are Enstatite chondrules, in order to give the mixture an overall non-carbonaceous nucleosynthetic isotope flavor.  
\begin{table}[h!]
Table 1: Element Abundances \& Condensation Temperatures \\
\begin{tabular}{llllll}
\toprule
Element                    &    Mg        &    Si         &     Na      &     K     &     Zn   \\
\midrule
CI Chondrites          &  95170   &  107740   &   5100    &   539   &    310   \\
Bulk Earth                &  151000   &  160000   &   2201    &    227   &    45.9   \\
Pebbles                  &  144000   &  167000   &   3970    &    483   &    135   \\
Bulk Mars             &  150000   &  174000   &   3600    &    300   &    89   \\   
\midrule
$T_{50}$  (K)  & 1336  &   1310  &  958   &   1006      &    726     \\
\bottomrule
\end{tabular}   
\noindent  

Abundances in ppm by mass. Bulk Earth from Wang et al. (2018); \\
CI chondrites from Lodders (2021); Pebble compositions from Garai et al. (2025a); \\
Bulk Mars from Yoshizaki and McDonough (2020);  50\% condensation temperatures from Lodders (2003). 
\end{table}

\begin{figure}[h!]
\begin{center}
\includegraphics[width=0.7\linewidth]{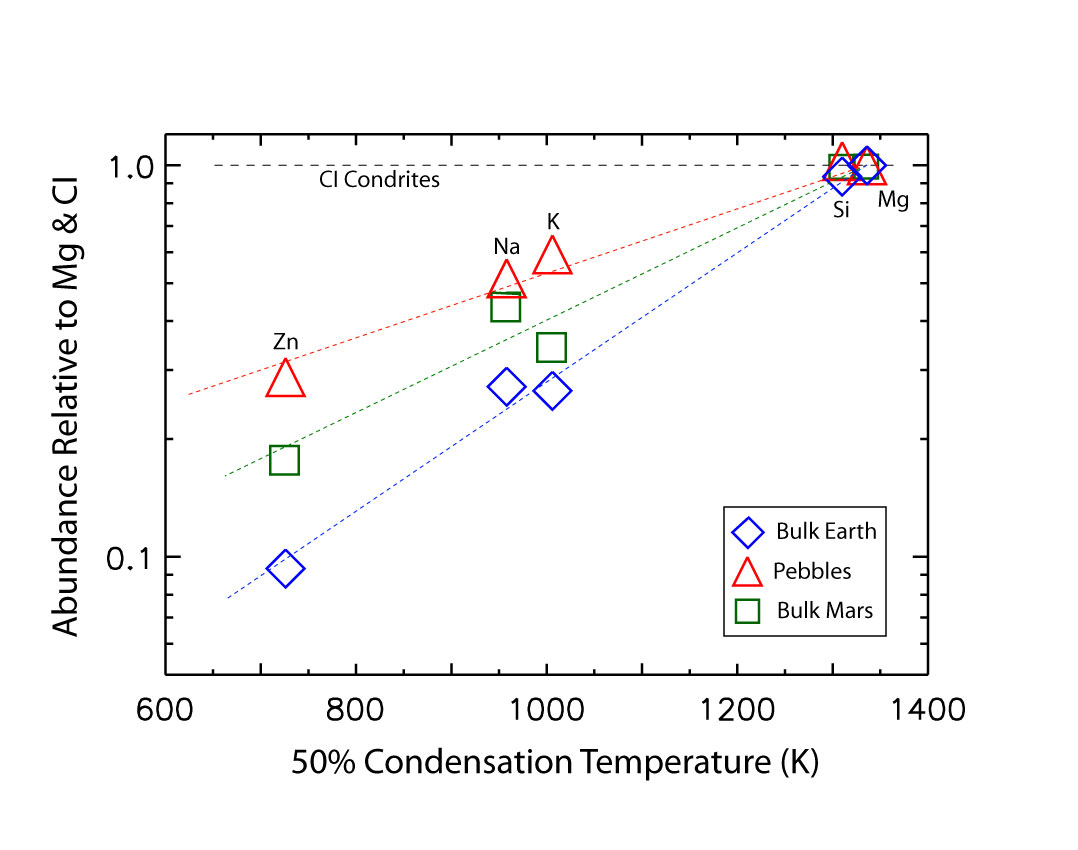}
\caption{Element abundances normalized by Mg and CI chondrite abundances versus condensation temperature. Bulk Earth (diamonds), Bulk Mars (squares), CI chondrite (dashed line), and Pebble (triangles) abundances and 50\% condensation temperatures are from Table 1.  }
\label{Dep3}
\end{center}
\end{figure}

Figure \ref{Dep3} shows the abundances of  Si, Na, K, and Zn versus their condensation temperature for the bulk Earth, bulk Mars, and our pebble model, relative to their abundance in CI chondrites. An additional normalization has been applied, such that the abundances of Mg in the Earth, Mars, and the pebbles match that in CI chondrites. This normalization scheme is commonly used to characterize volatile depletion by thermal processing in early Solar System materials and planets (Braukmüller et al., 2019; Bizzarro et al., 2025; McDonough, 2025). It is based on the dual premise that CI chondrites experienced minimal thermal processing and are therefore representative of primitive volatile-containing solids, and that Mg is a refractory element and experiences negligible loss from thermal processing.  

The relative abundances from our pebble model define a trend in Figure \ref{Dep3} that is intermediate between the volatile-rich CI chondrites and the more depleted bulk Earth. This suggests that Earth-building pebbles, unlike CI chondrites, may have experienced some thermal processing when they were formed. Nevertheless, they retained these moderately volatile elements in greater proportions than the bulk Earth. In this paper, we adopt this interpretation, and we use the pebble abundances listed in Table 1 for the input volatiles in our model.  Also shown in Figure \ref{Dep3} are the relative abundances of these same elements for bulk Mars. Note that the trend for bulk Mars lies between our pebble model and the bulk Earth, which suggests that depletion of moderately volatile elements might scale with protoplanet mass. The results presented below generally support this interpretation.

\section{Nebular gas flow near a growing protoplanet}
\begin{figure}[h!]
\begin{center}
\includegraphics[width=0.6\linewidth]{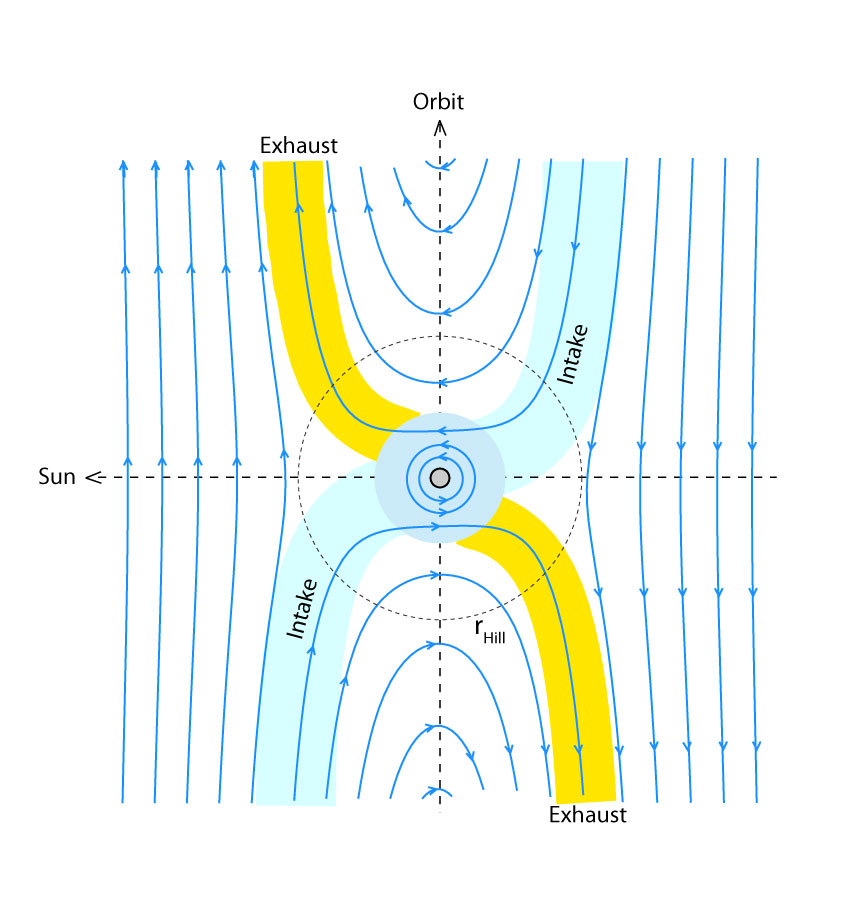}
\caption{Schematic of the gas circulation around a terrestrial protoplanet growing by pebble accretion. Blue contours are disk plane gas streamlines in a reference frame orbiting with the protoplanet. Dotted circle denotes the Hill radius, solid circle the protoplanet surface (exaggerated in size), circular blue shading denotes the protoplanet atmosphere. Labeled shadings denote pathways for intake of volatile-bearing pebbles and exhaust of thermally-processed volatiles. }
\label{Sketch1}
\end{center}
\end{figure}

Figure \ref{Sketch1} illustrates the primary two-dimensional circulation in the neighborhood of a pebble-accreting protoplanet embedded in a protoplanetary disk. Blue contours depict the time-average streamlines of gas motion in the plane of the disk, measured relative to the orbiting protoplanet. Blue shading denotes the protoplanet atmosphere (often called the {\it envelope}), where pressure, density, temperature, and gas residence time rise far above their far-field disk values. The solid and dotted circles denote the protoplanet surface and its Hill sphere, respectively. 

Two-dimensional models (Ormel, 2013; Bethune and Rafikov, 2019b) and three-dimensional models (Bethune and Rafikov, 2019a; Moldenhauer et al., 2021; 2022) reveal that this primary circulation includes three distinct regions. The first region is the background Keplerian shear flow, with streamlines sub-parallel to the orbital direction. The second region consists of nested, U-shaped streamlines symmetric about the protoplanet orbit, both fore and aft. The third region, the atmosphere/envelope, includes a strong prograde azimuthal wind with nearly circular (but slightly open) streamlines centered on the protoplanet. As shown in the Appendix, the intensity and prograde direction of this circulation is a consequence of amplification of the background (i.e., far-field) vorticity in the disk. In this study, we focus on how nebular gas, volatiles, and pebbles interact within the atmosphere region.

The other shadings in Figure \ref{Sketch1} mark the curved trajectories along which gas and pebbles enter the atmosphere, as well as the trajectories along which gas and vaporized species exit. These spiral arms -- one pair for intake, another pair for exhaust -- are found in both 2D and 3D circulation models. The biggest difference between 2D and 3D models are the secondary circulations found in the 3D models. This secondary circulation (not shown in Figure \ref{Sketch1}) is perpendicular to the disk plane and consists of radial inflows directed toward the protoplanet in both polar regions, and radial outflows directed away from the protoplanet at lower latitudes (Bethune and Rafikov, 2019a). 

One effect of the secondary circulation is to further ventilate the atmosphere by shortening the timescales for exhaust over what would be the case if the circulation were strictly 2D. In particular, exhaust timescales defined by $\tau_{exh}=M_{atm}/\dot{M}_{atm}$, where $M_{atm}$ is the atmosphere mass inside radius $r$ and $\dot{M}_{atm}$ is its rate of change, are found to be of order one hundred orbital periods in 3D isochemical and isothermal circulation models, even at $r$-values close to the protoplanet surface (Moldenhauer et al., 2022). In contrast, when the effects of gas cooling and volatile release are included, some model results indicate that the resulting stratification lengthens the exhaust timescales considerably, particularly close to the protoplanet (Kurokawa and Tanigawa, 2018). 

Despite the large differences between circulation models regarding exhaust timescales, there is a general consensus that highly volatile species evaporate high in the atmosphere and therefore exhaust quickly, on the order of a few orbital periods. This list includes volatile ices (Johansen et al., 2021), water (Wang et al., 2023), and sulfur (Steinmeyer et al., 2023). Similarly, there is general agreement that refractory silicates, which are abundant in vapor form only very close to the protoplanet surface, exhaust far more slowly, if at all (Steinmeyer and Johansen, 2024). 

The question addressed here is how moderately volatile elements such as Na, K, and Zn would fare under the same conditions.  These (and other) moderately volatile elements are less abundant then those listed above, yet they are key contributors to the volatile depletion trends observed for Earth (Li et al., 2021; Bizzarro et al., 2025; McDonough, 2025) and Mars (Yoshizaki and McDonough, 2020), as illustrated in Figure \ref{Dep3}. In addition, they are expected to be abundant in unheated pebbles, and to evaporate high enough in nebular atmospheres for exhaust processes to reduce their abundances substantially (Wang et al., 2025). Accordingly, they represent important test cases for pebble accretion. 

A widely-cited argument against pebble accretion is that it is unlikely to produce exponential depletion trends for modernly volatile elements, as implied by Figure \ref{Dep3}, because volatile depletion under pebble accretion is expected to be more step-like with respect to condensation temperature (Morbidelli et al., 2025), similar what is observed on Vesta (Sossi et al., 2022). In the following we show that volatile loss during pebble accretion can in fact yield an Earth-like depletion pattern, particularly if that protoplanet later merges with one or more less-depleted impactors, a likely outcome of planetary growth (Chambers, 2023).  

\section{Pebbles in a nebular atmosphere}
\begin{figure}[h!]
\begin{center}
\includegraphics[width=0.5\linewidth]{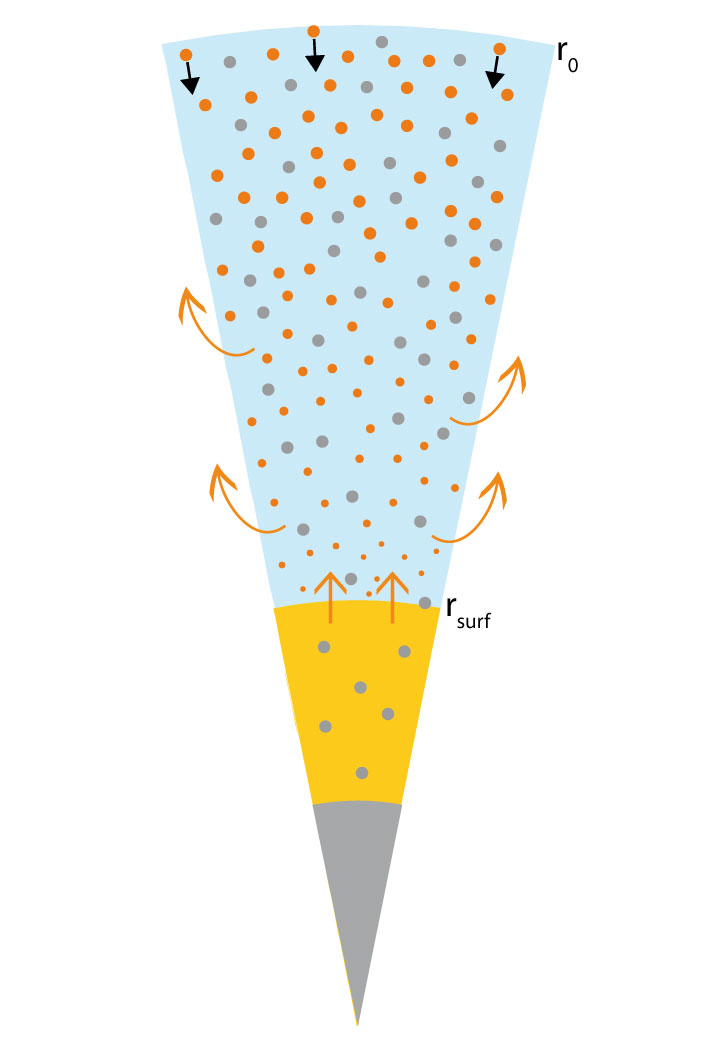}
\caption{Schematic of volatile fluxes into and out of a terrestrial protoplanet growing by pebble accretion. Blue, brown and gray shadings denote atmosphere, silicate mantle, and metallic core, respectively. Brown dots denote volatile-bearing silicate pebbles with size reduction indicating volatile loss to the atmosphere by evaporation. Gray dots denote metal pebbles, volatile-free in the atmosphere. Black arrows denote volatile addition by pebble influx. Short brown arrows denote volatile evaporation from the mantle; long curved brown arrows denote volatile loss from the atmosphere. }
\label{Sketch2}
\end{center}
\end{figure}

Figure \ref{Sketch2} illustrates the main interactions between pebbles, the atmosphere, and the protoplanet in our model. A mixture of volatile-bearing silicate pebbles and volatile-free metal pebbles settle through the atmosphere at terminal velocity. Both the metal and the silicate pebbles heat the atmosphere (as well as themselves) through release of their gravitational potential energy as they descend toward the protoplanet surface. Volatiles are added to the atmosphere by sublimation and evaporation of the silicate pebbles (depicted by their size reduction) and also by sublimation and evaporation from the protoplanet surface, particularly when the surface is hot enough to support magma ocean conditions. Volatiles are lost from the atmosphere along the exhaust trajectories shown in Figure \ref{Sketch1}. Volatiles are also removed from the mantle and transferred to the core via partitioning into sinking metal pebbles. 

There are multiple timescales implied in Figure \ref{Sketch2} that affect the final volatile content of the protoplanet. These include the accretion timescale $\tau_{acc}$, the atmosphere exhaust timescale $\tau_{exh}$,  the atmosphere saturation timescale $\tau_{sat}$, the time required to bring the atmosphere from zero concentration to saturation in a particular volatile species, the pebble settling timescale, i.e., their time-of-flight through the atmosphere  $\tau_{fly}$, and the pebble stopping time in the atmosphere $\tau_{stp}$. In addition, there are two timescales for vaporization, one for the pebbles, $\tau_{pev}$ and one for the mantle $\tau_{mev}$. Nominal values for these timescales are listed in Table 2, based on the characteristic sizes and densities of pebbles consisting of chondrules and metal grains accreting onto a protoplanet mass of $0.7M_{E}$, where $M_{E}$ is Earth's mass. 
\begin{table}[h!]
Table 2: Pebble Accretion Timescales \\
\begin{tabular}{llll}
\toprule
Timescale  &    Notation    &  Definition  &   Characteristic Value  \\
\midrule
Protoplanet Accretion  &   $\tau_{acc}$      &  $M/\dot{M_{p}}$              &  4 Myr     \\
Mantle Vaporization  &   $\tau_{mev}$      &  $M_{mv}/\dot{M}_{mv}$  &  4 Myr     \\
Atmosphere Exhaust   &   $\tau_{exh}$        &  $M_{atm}/\dot{M}_{atm}$      &   200 yr    \\
Atmosphere Vapor Saturation  &   $\tau_{sat}$    &  $x_{sat}/\dot{x}$    &    200 yr    \\
Pebble Vaporization  &   $\tau_{pev}$    &    $m_{pv}/\dot{m}_{pv}$    &    0.2 yr   \\
Pebble Flight Time   &     $\tau_{fly}$     &     $5r_{surf}/v_{p}$                 &    0.3 yr   \\
Pebble Stoping Time   &     $\tau_{stp}$     &     $\rho_{p}d_{p}^2/18\eta$      &    100 s   \\
\bottomrule
\end{tabular}   
\noindent
\end{table}

Our model is predicated on the following inequalities: $\tau_{mev}\simeq\tau_{acc} >> \tau_{exh} \simeq \tau_{sat} >> \tau_{pev}\simeq\tau_{fly} >> \tau_{stop}$, and also the assumption that the dynamical adjustment timescales in the atmosphere and magma ocean are short compared to $\tau_{acc}$. Taken together, these imply that pebble velocities are terminal, and the temperatures and volatile contents of the pebbles and the atmosphere are in an approximately steady-state balance at any given time, and that both are in approximate steady-state balance with the mantle in terms of temperature and volatile abundances. 

Note that Figure \ref{Sketch2} does not explicitly include azimuthal winds or turbulence.  These become major effects in situations where the pebble size is nonuniform, by virtue of heterogeneous conditions at their formation, through fragmentation and aggregation within the atmosphere, or both (Popovas et al., 2018; Johansen and Nordlund, 2020). In order to minimize the complications these factors produce, we assign a uniform pebble Stokes number and a uniform settling velocity at every altitude, implying no effects from pebble fragmentation or aggregation. This is not actually correct because silicate and metal pebbles have different densities and also because the silicate pebbles partially vaporize, but the errors it produces are not significant for our purposes.  We also assign a uniform turbulent thermal diffusivity to the atmosphere, which we use as a modeling parameter in calculating the atmosphere temperature profile, but we ignore the effects of turbulence on pebble settling rates.  Atmospheric particle studies (e.g., N\"{a}slund and Thaning, 1991) typically find only minor ($\sim$10\%) increases in the average settling time from turbulence for particles in the Reynolds number range of our pebbles. 

Because the timescales of accretion and mantle vapor loss are commensurate, we must include the kinetics of vapor loss from the mantle. We use the overdot notation for time derivatives on these long timescales. Similarly, because the timescales of pebble flight and pebble vaporization are commensurate, we must also include pebble vaporization kinetics.  We use $d/dt$ notation for time derivatives on these short timescales. Lastly, because all of the effects of compositional stratification on atmosphere exhaust have yet to be fully explored, in this study we treat the atmosphere exhaust timescale as modeling parameter. We use a nominal value of $\tau_{exh}$=200 yr for the whole atmosphere in our reference cases, and then later we examine the sensitivity of our results to wide variations of this timescale. 

\section{Model formulation}

\subsection{Pebble types}
Pebbles arriving at the top of the atmosphere in our model are a homogeneous mixture consisting of two types, hereafter referred to as silicate and metal, respectively. The metal pebbles comprise a fixed fraction $\mu$ of the total pebble mass, the silicate pebbles comprising the remainder. For purposes of simplicity, we assign all of the volatiles to the silicate pebbles. The metal pebbles and the remaining mass fraction of the silicate pebbles are treated as volatile-free. The next subsection describes how we track volatile release from silicate pebbles, their mass loss, and their settling rates. In contrast, the metal pebbles conserve their individual mass and their total mass flux during settling. Prior to volatile loss, the silicate pebbles have a nominal 1 mm diameter and 3$\times 10^3$ kg/m$^3$ mean density. However, we assign a higher mean density to silicates in the protoplanet mantle, in order that the protoplanet radius be realistic for a terrestrial body under hydrostatic compression.  

\subsection{Pebble flight and vaporization} 
Consistent with the assumptions and timescale restrictions listed in the previous section, the equations of motion for a small pebble flying through a nebular atmosphere include its settling rate
\begin{equation}
\frac{dy_{p}}{dt} = -v_{p},
\label{palt}
\end{equation}
where $y_{p}$ is the altitude of the pebble above the protoplanet surface, $t$ is time of flight, and the (downward) terminal speed of the pebble is
\begin{equation}
v_p = \big(\frac{m_{p}^{1/3}\rho_{p}^{2/3} g}{\rho A_{p}C_{d}}\big)^{1/2}. 
\label{pvel}
\end{equation} 
Here $m_{p}$ and $\rho_{p}$ are pebble mass and density, respectively, $g$ is gravity, $\rho$ is atmosphere density, $A_{p}$ is the pebble shape factor, and $C_{d}$ is the
pebble drag coefficient, given by
\begin{equation}
C_{d} = {\rm Ma} + \frac{12}{\rm{Re}} + \frac{2}{{\rm Re}^{1/2}} + 1,        
\label{pdrag}
\end{equation}
where ${\rm Re}$ and ${\rm Ma}$ are the pebble Reynolds and (modified) Mach numbers, defined as 
\begin{equation}
{\rm Re} = \frac{\rho d_{p} v_{p}}{\eta}; \;\;\;\; {\rm Ma}=\frac{v_p}{\pi c}. 
\label{pRe}
\end{equation}
The first term in (\ref{pdrag}) represents Epstein drag, the second represents Stokes drag, the fourth represents quadratic drag, and the third
term represents the transition between the second and fourth. In (\ref{pRe}), $d_{p}$ is the pebble diameter, $\eta$ is the local atmosphere viscosity, and $c$ is the local sound speed defined below in terms of atmosphere properties. 

The heat balance for the same pebble can be written as (Adolfsson et al., 1996)
\begin{equation}
\frac{dT_{p}}{dt} = \frac{A_{p}} {C_{p}m_{p}^{1/3}\rho_{p}^{2/3}} \bigg(\frac{\rho C_{h}v_{p}^3}{2}-4k_{s}\epsilon(T_{p}^4-T^4)
   + \frac{L_{p}\rho_{p}^{2/3}}{A_{p}m_{p}^{2/3}}\frac{dm_{pv}}{dt} \bigg),
\label{ptemp}
\end{equation}
where $T_{p}$ is the average temperature of the pebble, $T$ is the local temperature of the atmosphere, $C_{p}$ is pebble specific heat, $C_{h}$ is the pebble heat transfer coefficient, $k_{s}$ is the Stefan-Boltzmann constant, $\epsilon$ is emissivity, $L_{p}$ is pebble heat of vaporization, and $m_{pv}$ is the mass of volatiles in the pebble. The terms on the r.h.s. of (\ref{ptemp}) represent frictional heating, radiative heat loss, and latent heat loss, respectively. 

For calculating the vaporization rate of silicate pebbles, we use Hertz-Knudsen kinetics (Hill et al., 2005), in which volatile mass evolves according to
\begin{equation}
\frac{dm_{pv}}{dt} = -4A_{p}(\frac{m_{p}}{\rho_{p}})^{2/3} (\frac{{\rm m}_{v}} {2\pi R T_{p}})^{1/2}(P_{pev}-P_{v}),
\label{pvap}
\end{equation}
where ${\rm m}_{v}$ is the molecular weight of the volatile in kg/mol, $R$ is the universal gas constant, $P_{v}$ is the local atmosphere vapor pressure,
and $P_{pev}$ is the equilibrium vapor pressure for the pebble volatile. We apply ({\ref{pvap}) separately to each volatile species and track the changes in pebble mass relative to $m_{p0}$ and $m_{pv0}$, the pebble mass and its volatile mass at the top of the atmosphere.  The equilibrium vapor pressure for each volatile species is calculated relative to a standard composition as follows:
\begin{equation}
\log_{10} P_{pev} = (\gamma_{1} + \log{x^*_{p}}) - \frac{\gamma_{2}}{T_{p}}, 
\label{peqvp}
\end{equation}
where $\gamma_{1}$ and $\gamma_{2}$ are experimentally-determined coefficients and $x^*_{p}=x_{p}/x_{std}$  is the Raoult's law factor correcting for the difference between $x_{p}$, the volatile concentration in the pebble and the standard
concentration of that volatile, $x_{std}$. The formula used for the atmosphere vapor pressures $P_{v}$ is given in the next subsection. Units and values of the variables and properties in (\ref{pvel})-(\ref{peqvp}) are given in Tables 3-6. In particular, Table 6 gives the values of the $\gamma$-coefficients and the volatile concentrations we use as standards. These were derived from vaporization determinations on ordinary chondrites for SiO and Na by Dias et al. (2020), and on basaltic Eucrites for K and Zn by Kitts and Lodders (1998) and Tian et al. (2019). 
\clearpage
\newpage
\begin{table}[h!]
Table 3: Pebble Accretion Variables   \\
\begin{tabular}{lll}
\toprule
Variable    & Notation &  Units  \\
\midrule
Planet mass, accretion rate & $M$, $\dot{M}$  & $kg$, $kg/s$   \\
Mantle mass, mantle volatile mass       &     $M_{m}, M_{mv}$       &  $kg$   \\
Pebble mass, volatile mass & $m_{p}$, $m_{pv}$  & $kg$   \\
Atmosphere pressure & $P$  & $Pa$   \\
Atmosphere vapor pressure & $P_{v}$  & $Pa$   \\
Radial distance & $r$   & $m$   \\
Cylindrical coordinates, disk-normal & $s,\phi,z$  &   $m,rad,m$   \\
Atmosphere temperature & $T$  & $K$   \\
Pebble temperature & $T_{p}$  & $K$   \\
Mantle surface temperature   &    $T_{surf}$     &           $K$       \\
Time & $t$  & $s$   \\
Pebble settling speed & $v_{p}$  & $m/s$   \\
Atmosphere azimuthal wind   &    $v_{\phi}$       &      $m/s$          \\
Atmosphere volatile concentration, molar   &    $X$       &     $mol/mol$      \\
Mantle volatile concentration, molar   &    $X_{m}$       &     $mol/mol$      \\
Pebble volatile concentration, molar   &    $X_{p}$       &     $mol/mol$      \\
Atmosphere volatile concentration, wt.   &    $x$       &     $kg/kg$      \\
Mantle volatile concentration, wt.   &    $x_{m}$       &     $kg/kg$      \\
Pebble volatile concentration, wt.   &    $x_{p}$       &     $kg/kg$      \\
Pebble altitude & $y_{p}$  & $m$   \\
Atmosphere potential vorticity &  $\Gamma$      &        $m^2$/$kg$-$s$         \\
Atmosphere density & $\rho$  & $kg/m^{3}$   \\
Atmosphere vorticity &  $\omega$      &        $rad/s$         \\
\bottomrule
\end{tabular}
\end{table}

\clearpage
\newpage
 \begin{table}[h!]
Table 4: Pebble Accretion Variable Properties  \\
\begin{tabular}{lll}
\toprule
 Variable Property    & Notation &  Units  \\
\midrule
Pebble drag coefficient  & $C_{d}$  &  $n.d.$  \\
Atmosphere specific heat   &   $C_{P}$       &    $J/kg$-$K$  \\
Atmosphere sound speed & $c$  &  $m/s$    \\
Pebble diameter & $d_{p}$  & $m$   \\
Atmosphere total heat flux    &  $F_{atm}$       &     $W$       \\
Pebble sensible heat flux    &  $F_{sen}$       &     $W$       \\
Atmosphere gravity & $g$  & $m/s^2$   \\
Surface gravity & $g_{surf}$  & $m/s^2$   \\
Atmosphere mass     &     $M_{atm}$       &  $kg$   \\
Pebble Mach, Reynolds numbers    &     ${\rm Ma}$, ${\rm Re}$    &   $n.d.$   \\
Core mass  &   $M_{c}$      &    $kg$          \\
Volatile influx, pebbles    &    $\dot{M}_{pv}$       &     $kg/s$       \\
Volatile vaporization rate, mantle    &    $\dot{M}_{mve}$       &     $kg/s$       \\
Volatile precipitation rate, pebbles    &    $\dot{M}_{pvm}$       &     $kg/s$       \\
Volatile vaporization rate, pebbles    &    $\dot{M}_{pve}$       &     $kg/s$       \\
Volatile exhaust rate, atmosphere    &    $\dot{M}_{vex}$       &     $kg/s$       \\
Atmosphere mean molecular weight  &   $\bar{\rm m}$    &   $kg/mol$  \\
Pebble mass/starting mass & $m^*_{p}$   & $n.d.$   \\
Atmosphere stability parameter    &     $N^{2}$        &       $rad^2/s^2$   \\
Vapor pressure, mantle equilibrium & $P_{mev}$  & $Pa$   \\
Vapor pressure, pebble equilibrium & $P_{pev}$  & $Pa$   \\
Atmosphere pressure, starting   &      $P_{0}$       &      $Pa$    \\
Pebble heat production  & $Q_{p}$  & $W/m^3$   \\
Atmosphere turbulent, radiative heat fluxes & $q_{turb}$, $q_{rad}$  & $W/m^2$   \\
Atmosphere gas constant  &    $R'$    &     $J/kg$-$K$    \\
Atmosphere vapor mixing ratio & $R_{mix}$  & $kg/kg$   \\
Atmosphere starting, surface radii     &        $r_{0}$, $r_{surf}$          &      $m$      \\
Mantle volatile concentration, starting   &    $x_{m0}$       &     $kg/kg$      \\
Atmosphere volatile saturation concentration, wt.   &    $x_{sat}$       &     $kg/kg$      \\
Raoult's law factors, pebbles, mantle     &        $x^*_{p}, x^*_{m} $   &     $n.d.$          \\
Atmosphere viscosity & $\eta$  & $Pa$-$s$   \\
Pebble density & $\rho_{p}$  & $kg/m^3$   \\
Atmosphere gas column density    &     $\Sigma$      &   $kg/m^2$    \\
Atmosphere opacity & $\sigma$  & $m^2/kg$   \\
Atmosphere vapor pressure fraction   &  $\chi$     &   $n.d.$         \\
 \bottomrule
\end{tabular}
\end{table}

\clearpage
\newpage
\begin{table}[h!]
Table 5: Pebble Accretion Constants and Constant Properties  \\
\begin{tabular}{lll}
\toprule
Constant    & Notation &  Value  \\
\midrule
Gravitational  &   $G$     &   6.67$\times 10 ^{-11}$ $m^3/kg$-$s^2$   \\
Stefan-Boltzmann   &    $ k_{s} $    &   5.67$\times 10^{-8}$  $W/m^2$-$K^4$  \\
Universal gas         &         $R$          &     8.314 $J/K/mol$     \\
Year, Day             &    $yr$, $day$        &    $3.156\times 10^7$, $8.64\times 10^4$   $s$        \\
\midrule
Property    & Notation &  Value(s)  \\
\midrule
Pebble shape factor         &           $A_{p}$     &         1.2   $n.d.$  \\
Pebble heat transfer coefficient    &    $C_{h}$     &      1   $n.d.$  \\
Pebble specific heat    &    $C_{p}$     &      1.25$\times 10^3$  $J/kg$-$K$     \\
Metal-silicate partition coefficient     &    $D_{cm}$     &    Table 6         \\
Atmosphere basic mixture    &    H$_{2}$, He    &    0.854, 0.146  $mol/mol$ \\
Pebble latent heat, melting \& vaporization   &    $L_{p}$     &   6$\times10^6$  $J/kg$     \\
Earth mass, present	  &   $M_{E}$      &  5.97$\times 10^{24}$   $kg$ \\     
Solar mass    &  $M_{S}$      &  1.99$\times 10^{30}$   $kg$ \\
Pebble accretion rate          &   $\dot{M}_{p}$    &    0.2 $M_{E}/Myr$   \\
Silicate pebble volatile molecular weight  &   ${\rm m}_{v}$    &   Table 6 \\
Nebula pressure     &   $P_{n}$     &     0.1  $Pa$    \\
Nebula temperature     &   $T_{n}$     &     150  $K$    \\
Pebble rheological transition temperature      &   $T_{rt}$      &      1700 K    \\
Atmosphere temperature, starting       &  $T_{0}$     &        200 $K$     \\
Condensation temperature            &       $T_{50}$     &        Table 1    \\
Pebble mass, volatile mass, starting & $m_{p0}$, $m_{pv0}$  & various $kg$    \\
Proto-Earth orbital radius  &   $r_{\Omega}$   &   1.5$\times 10 ^{11}$ $m$ \\
Silicate pebble volatile concentration, starting  &    $x_{p0}$     &     Table 6 \\
Silicate volatile concentration, standard  &    $x_{std}$     &     Table 6 \\
Equilibrium vapor pressure coefficients, standard       &       $\gamma_1, \gamma_2$    &     Table 6  \\
Pebble emissivity    &         $\epsilon$          &      0.9     $n.d.$  \\
Nebula gas column density    &     $\Sigma_{n}$     &     1$\times 10^3$   $kg/m^2$    \\   
Atmosphere turbulent diffusivity   &     $\kappa_{turb}$      &      $10^{4}$      $m^2/s$    \\
Metal pebble fraction      &      $\mu$        &    0.325  $kg/kg$      \\     
Core-forming metal density           &    $\rho_{c}$  &   10$\times 10^3$    $kg/m^3$ \\
Mantle-forming silicate density     &   $\rho_{m}$  &   4$\times 10 ^3$    $kg/m^3$ \\
Atmosphere opacity parameters, H$_2$-He   &    $\sigma_2, a_2, b_2$     &  1$\times 10^{-8}$, 0.667, 3.0     \\  
Atmosphere opacity parameters, dust   &    $\sigma_1, a_1, b_1$     &  2$\times 10^{-3}$, 0, 0.5     \\   
Proto-Earth orbital velocity  &   $\Omega$   &   2$\times 10 ^{-7}$ $rad/s$ \\
\bottomrule
\end{tabular}    
\end{table}

\clearpage    
\newpage
\begin{table}[h!]
Table 6: Volatile Properties \\
\begin{tabular}{lllll}
\toprule
Volatile   &  ${\rm m}_{v}$         & Concentration (Bulk Earth,  &  \; $\gamma_1$ \;\;\;\;\; $\gamma_2$  &  $D_{cm}$   \\
           &                                     &         Silicate Pebbles,            &                                                              &                    \\
           &                                     &      Standard Composition)      &                                                              &                    \\
\midrule
SiO   &  44     &                    (0.25, 0.388, 0.35)                  &    13.0 \;\; 2.88$\times 10^4$       & 0.023   \\
Na   &  23    &               (2.2, 5.88, 3.8)$\times 10^{-3}$     &    8.67 \;\; 1.44$\times 10^4$      & 0.527   \\
K    &  39    &               (2.27, 7.15, 3.3)$\times 10^{-4}$    &    7.97 \;\;  1.68$\times 10^4$    & 0.89   \\
Zn  &  65    &                 (0.46, 2.0, 0.2) $\times 10^{-4}$   &    6.69 \;\;   1.09$\times 10^4$     &  0.53    \\
\bottomrule
\end{tabular}   
\noindent  

   Molecular weights ${\rm m}_{v}$  in kg/kmol. Concentrations in kg/kg. Bulk Earth from Wang et al. (2018); 
   Silicate pebbles from Garai et al. (2025a).  Standard composition and  $\gamma$-coefficients
   (defined in text) from Dias et al. (2020) for SiO and Na;  Tian et al. (2019) and Kitts and Lodders (1998) 
   for K and Zn. Core/mantle partition ratios $D_{cm}$  (defined in text) from Wang et al. (2018). 
\end{table}

\subsection{Atmosphere structure}
Based on the timescale inequalities listed in the previous section, we consider a 1D hydrostatic atmosphere heated by settling pebbles. The basic dry (i.e., volatile-free) atmosphere
has a solar-like composition (0.854H$_{2}$, 0.146He), and consists of an optically thin, nearly isothermal outer layer and an optically thick inner layer with a variable thermal gradient. Volatile addition occurs
in the inner layer only. 

The governing equations for the inner layer include the hydrostatic balance 
\begin{equation}
\frac{dP}{dr} = -\rho g = -\frac{\rho GM}{r^2}, 
\label{ahydro}
\end{equation}
where $P$ and $\rho$ are gas hydrostatic pressure and density, $G$ is the gravitational constant, $M$ is protoplanet mass (the atmosphere mass being negligible as far as gravity is concerned), 
and $r$ is radial distance from the protoplanet center, the ideal gas law
\begin{equation}
P = \rho R' T,
\label{ideal}
\end{equation}
with a modified gas constant given by $R'=R/ \bar{\rm m}$,
where $\bar{\rm m}$ is the mean atomic weight of the atmosphere gas in kg/mol.  For the
local sound velocity we use $c^2=1.42R'T$.

For heat transport in the inner atmosphere, we modify the standard equilibrium heat balance for a radiative-turbulent atmosphere (Liou and Ou, 1983) as follows:
\begin{equation}
\frac{\partial}{\partial t} (\rho C_{P}T) + \nabla_{r} (q_{turb} + q_{rad}) = Q_{p}    
\label{heat}
\end{equation}
where $\nabla_{r} \equiv d/dr + 2/r$ and $C_{P} = 3.38R'$ is atmosphere specific heat, 
\begin{equation}
q_{turb} = -\rho C_{P} \kappa_{turb} (\frac{dT}{dr} + \frac{g}{C_{P}} + \frac{L_{p}}{C_{P}}\frac{dR_{mix}}{dr}) 
\label{turb}
\end{equation}
is the local turbulent heat flux,
\begin{equation}
q_{rad} = -\frac{16k_{s}T^3}{3\rho\sigma}\frac{dT}{dr}
\label{rad}
\end{equation}
is the local radiative heat flux, and
\begin{equation}
Q_{p} = \frac{\dot{M}_{p}}{4\pi r^2} m^*_{p} (g + C_{p}\frac{dT_{p}}{dr} - \frac{L_{p}}{m_{p}}\frac{dm_{p}}{dr}) 
\label{Qp}
\end{equation}
is the (net) heat source from the settling pebbles. In (\ref{turb})-(\ref{Qp}), $R_{mix}$ is the (saturation) mixing ratio of the volatile species, $\kappa_{turb}$ is the turbulent diffusivity of heat, $\dot{M}_{p}$ is the pebble accretion rate at the top of the atmosphere, $m^*_{p}=m_{p}/m_{p0}$, and $\sigma$ is atmosphere opacity. The bracketed terms in (\ref{Qp}) represent, respectively, the heat added to the atmosphere along with the sensible and latent heat removed from the atmosphere by falling pebbles.  
                                                                                                                                                                                                                   
Thermal equilibrium in the atmosphere at each stage of protoplanet evolution dictates that we set $\partial/\partial t=0$ in equation 
(\ref{heat}). Integrating (\ref{heat}) in radius from $r$ to $r_{0}$ and substituting (\ref{turb})-(\ref{Qp}) yields the following equation for atmosphere temperature: 
\begin{equation}
 \rho C_{P} \kappa_{turb} (\frac{dT}{dr} + \frac{g}{C_{P}} + \frac{L_{p}}{C_{P}}\frac{dR_{mix}}{dr})
 + \frac{16k_{s}T^3}{3\rho\sigma}\frac{dT}{dr} 
 = \frac{\dot{M}_{p}}{4\pi r^2}  \int_{r}^{r_{0}} m^*_{p} (g + C_{p}\frac{dT_{p}}{dr'} - \frac{L_{p}}{m_{p}}\frac{dm_{p}}{dr'})dr'  - q_{rad}(r_{0}),
\label{temp}
\end{equation}
in which we have assumed that $q_{turb}$ is negligible at $r_{0}$.  

The coupling between the atmosphere, the pebbles, and the protoplanet involves several variable
properties, including the atmosphere opacity and viscosity, along with the mixing ratio and vapor pressure of each volatile species in the atmosphere. 
The opacity function we adopt uses the temperature dependence developed by Bell and Lin (1994):
\begin{equation}
\sigma = \sum_{j} \sigma_{j}\rho^{a_j} T^{b_j}.
\label{opat}
\end{equation}
We include opacity contributions from hydrogen and helium gas molecules and sparse dust, with coefficients $\sigma_{j}$, $a_{j}$, and $b_{j}$ given in Table 5. 

For the atmosphere viscosity, we combine previously-determined viscosity versus temperature profiles for H$_2$, He, and SiO vapor, using the formulation for gas mixtures described in the Appendix. The individual viscosity versus temperature profiles for each of these species are shown in Figure \ref{Visc} along with the viscosity profile for our dry (vapor-free) H$_2$-He mixture. Note that the presence of He substantially increases the mixture viscosity. In contrast, including a small amount of SiO vapor changes the basic atmosphere viscosity profile by a few percent at most. The influence of the other volatiles on atmosphere viscosity is even smaller and is ignored. 
\begin{figure}[h!]
\begin{center}
\includegraphics[width=0.6\linewidth]{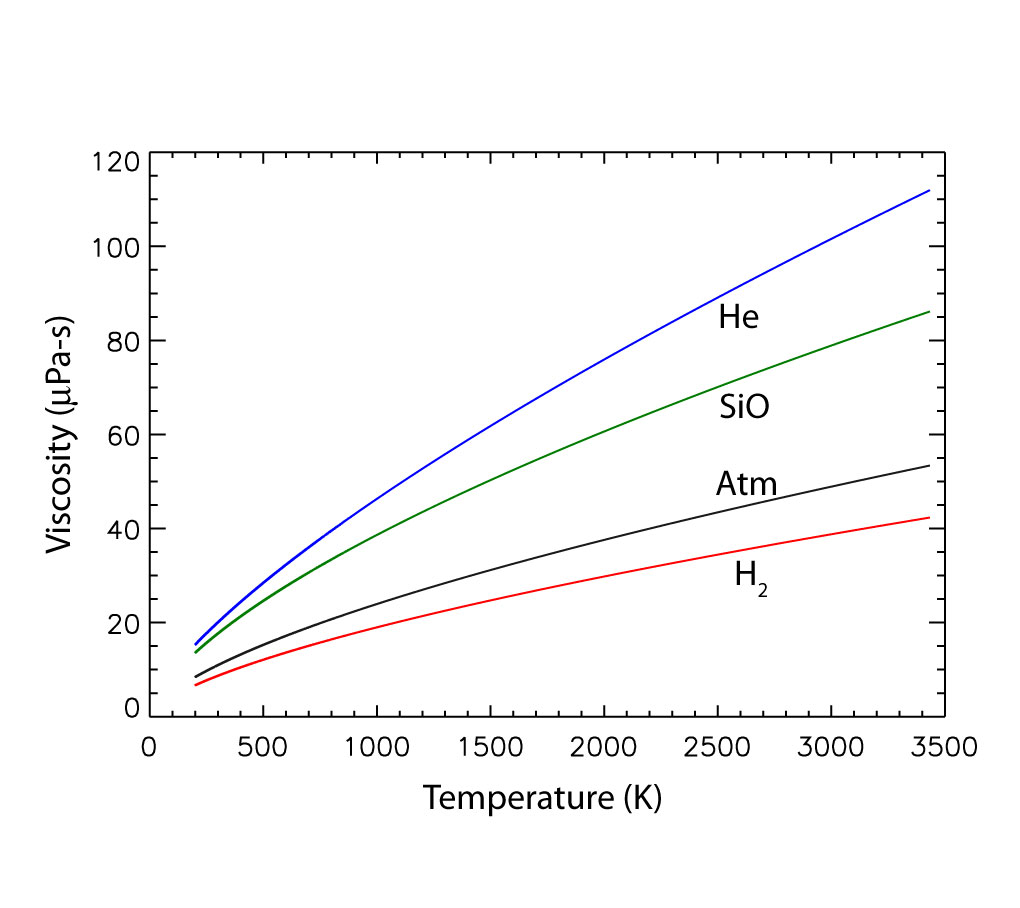}
\caption{Profiles of viscosity versus temperature for H$_{2}$, He, SiO vapor, and our volatile-free model H$_{2}$-He atmosphere.}
\label{Visc}
\end{center}
\end{figure}

One of our key assumptions is that the partial vapor pressures in the atmosphere are controlled by the local atmosphere temperature and the volatile abundances in the silicate mantle of the protoplanet. This is justifiable in the lowermost atmosphere, where most of the volatiles are concentrated. In particular, we assume that at each stage of protoplanet growth the atmosphere partial pressure of each volatile $P_{v}$ is some fraction $\chi$ of the equilibrium vapor pressure $P_{mev}$ that corresponds to the mantle abundance of that volatile:  
\begin{equation}
P_{v}= \chi P_{mev}.
\label{avp}
\end{equation}
The equilibrium vapor pressures in the atmosphere $P_{mev}$ are determined at each stage of protoplanet growth as described in the next subsection. The fractions $\chi$ in (\ref{avp}) are determined at each stage of growth using mass balance considerations, as described in the subsection after next. We find that $\chi$ is very close to one in all cases, indicating that the atmosphere remains very nearly volatile-saturated with respect to the mantle. 

The mixing ratio for each volatile species is given by $R_{mix} = x/(1-x)$, 
where $x$ is mass concentration of that volatile in the atmosphere, related to its molar concentration $X$ and molecular weight ${\rm m}$ via  
$x={\rm m}X/\bar{\rm m}$. The molar concentration of each volatile is the ratio of its partial pressure to the local atmosphere pressure, $X=P_{v}/P$. Note that, because $\chi \simeq$1, $R_{mix}$ is very nearly equal to the saturation mixing ratio. 

Several useful global diagnostics can be calculated for this atmosphere model.  These include the total atmosphere heat flux (in W), given by 
\begin{equation}
F_{atm}=4\pi r^{2}(q_{turb} + q_{rad}),
\label{Fatm}
\end{equation}
and the total sensible heat flux deposited onto the protoplanet surface by settling pebbles (also in W), given by
\begin{equation}
F_{sen} = \dot{M}_{p}C_{p}T_{surf}\big( (1-\mu)m^*_{p}(r_{surf}) + \mu\big).
\label{sen}
\end{equation}
In equation (\ref{sen}),  $r_{surf}$ is the surface radius, $m^*_{p}(r_{surf})$ is $m^*_{p}$ evaluated there, and we have equated the pebble and the atmosphere surface temperatures. 
Similarly, the total volatile mass flux added to the protoplanet mantle by pebble precipitation is
\begin{equation}
\dot{M}_{pvm} = \dot{M}_{p} (1-\mu)\big(1-m^*_{p}(r_{surf})\big) ,
\label{Mpvm}
\end{equation}
while the total mass flux of vapor exhausted from the atmosphere is given by 
\begin{equation}
\dot{M}_{vex}=M_{vex}/\tau_{exh}=\frac{4\pi}{\tau_{exh}}\int_{r_{surf}}^{r_0} x\rho r^2 dr'.
\label{Mvex}
\end{equation}
Lastly, we measure the stability of the atmosphere in terms of the squared buoyancy frequency, defined as  
\begin{equation}
N^2 = -(\frac{g}{\rho}\frac{d\rho}{dr} + \frac{g^2}{c^2}),
\label{Nsq}
\end{equation}
with positive values corresponding to stable stratification and negative values corresponding to unstable stratification.

\subsection{Protoplanet structure}
Our model protoplanet consists of a uniform silicate mantle with density $\rho_{m}$, a uniform metallic core with density $\rho_{c}$, and a time-dependent total mass $M$. In order to correct for mass lost by atmosphere exhaust, we equate the growth rate of the protoplanet to the mass influx of pebbles minus the vapor exhaust rate, i.e., we set $\dot{M}=\dot{M}_{p} - \dot{M}_{vex}$. Furthermore, we fix the relative masses of the mantle $M_{m}$ and the core $M_{c}$, such that $M_{c}/M=\mu$ and $M_{m}/M=1-\mu$, with $\mu$ constant at all stages of protoplanet growth. The mantle and core masses then evolve according to
\begin{equation}
\dot{M}_{c}=\mu \dot{M}_{p};    \;\;\;\;\;  \dot{M}_{m}=(1-\mu) \dot{M}_{p} - \dot{M}_{vex},
\label{Mcm} 
\end{equation}
and the radius and surface gravity of the protoplanet are given by
\begin{equation}
r_{surf} = \big(\frac{3(1-\mu)M}{4\pi\rho_{m}}   +   \frac{3\mu M} {4\pi\rho_{c}}\big)^{1/3}
\label{rsurf} 
\end{equation}
and
\begin{equation}
g_{surf} = \frac{GM}{r_{surf}^2}.
\label{gsurf} 
\end{equation}
We use the compressed (high pressure) values for $\rho_{m}$ and $\rho_{c}$ from Table 5 and $\mu$=0.325 in order to give Earth-like values for surface radius a gravity at one Earth mass. 

We set the surface concentration of each volatile equal to its mantle average concentration, and we set the surface temperature equal to the temperature at the base of the atmosphere. We apply the same criteria for determining the mantle equilibrium vapor pressures as we have for pebbles, namely
\begin{equation}
\log_{10} P_{mev} = (\gamma_{1} + \log_{10}{x^*_{m}}) - \frac{\gamma_{2}}{T},
\label{peqvp2}
\end{equation}
where $x^*_{m}=x_{m}/x_{std}$ is the mantle Raoult's law factor, the ratio of the volatile concentration in the mantle $x_{m}=M_{mv}/M_{m}$ to the standard volatile concentration. 
The actual vapor pressure for each volatile at and above protoplanet surface is given by equations (\ref{avp}) and (\ref{peqvp2}).  

For calculating volatile loss from the protoplanet, we treat the entire mantle as a giant pebble.
By analogy with equation (\ref{pvap}), the rate of volatile loss from the mantle is given by 
\begin{equation}
\dot{M}_{mve} = -4A_{p}(\frac{M_{m}}{\rho_{m}})^{2/3} (\frac{{\rm m}_{v}} {2\pi R T_{surf}})^{1/2}(1-\chi)P_{mev}.
\label{pvapm}
\end{equation}
In models that include protoplanet growth, we need to specify the initial mantle volatile concentrations at the start of growth. All of the evolutionary calculations presented here begin at a protoplanet mass of 0.2$M_{E}$.  For masses smaller than this, there is negligible loss of the four volatiles we consider. Accordingly, we assume zero volatile loss in the initial mass, so that the initial mantle volatile concentrations are given by 
\begin{equation}
x_{m0} = \frac{ (1-\mu)x_{p0}}{ \mu(D_{cm}-1)+1 },
\label{xmi}
\end{equation}
where $x_{p0}$ is the input pebble concentration and $D_{cm}$ is the core-mantle partition coefficient for each of the four volatiles in Table 6.  

\subsection{Volatile mass balances}
On the basis of the timescale inequalities in Section 4, we assume that the pebbles, the atmosphere, and the mantle are in an approximately statistically steady
state balance at each stage of protoplanet growth. This implies that the volatile mass flux carried into the atmosphere by pebbles equals 
the evaporation rate from pebbles plus the rate of volatile precipitation from pebbles settling onto the protoplanet surface (in the notation of this subsection we do not distinguish between evaporation and sublimation). 
Accordingly, for each volatile
\begin{equation}
\dot{M}_{pve} = \dot{M}_{pv} - \dot{M}_{pvm},
\label{PvB}
\end{equation}
where $\dot{M}_{pv}=x_{p0}\dot{M}_{p}$ and $\dot{M}_{pve}$ denote pebble volatile influx at the top of the atmosphere and pebble volatile evaporation, respectively, and $\dot{M}_{pvm}$ is given by equation (\ref{Mpvm}). The corresponding atmosphere volatile balance equates sum of the pebble influx $\dot{M}_{pv}$ plus the mantle volatile evaporation rate $\dot{M}_{mve}$ from (\ref{pvapm}) to the sum of the precipitation rate into the mantle $\dot{M}_{pvm}$ plus the volatile exhaust rate $\dot{M}_{vex}$ from equation (\ref{Mvex}). This balance yields      
\begin{equation}
\dot{M}_{mve} = \dot{M}_{vex} + \dot{M}_{pvm}  - \dot{M}_{pv}.
\label{AvB}
\end{equation}
Finally, the mantle volatile balance states that the change in its volatile content equals volatile precipitation minus the sum of evaporation and partitioning into the core:
\begin{equation}
\dot{M}_{mv} = \dot{M}_{pvm} - \dot{M}_{mve} - \mu D_{cm} \dot{M}_{p} m_{p}^*(r_{surf})  x_{m},
\label{MvB}
\end{equation}
where  $D_{cm}$ is the volatile core/mantle partition coefficient. 

\section{Solution procedure}
We solve the system of equations in the previous section using the following approach. For a given protoplanet mass, we determine the protoplanet structure using equations (\ref{Mcm})-(\ref{gsurf}). We then 
insert an outer, isothermal atmosphere layer that connects the disk pressure, temperature, and density to the pressure, temperature, and density at the top of the inner atmosphere layer, as described in the Appendix. We also specify an initial estimate of the atmosphere saturation factor $\chi$ for SiO.  

We then divide the pebble flux into silicate and metal parts based on the ratio $\mu$, and track the flight and thermal processing of an SiO-containing silicate pebble using equations (\ref{palt})-(\ref{peqvp}), starting at the top of the inner atmosphere at the altitude corresponding to $r_{0}$=10$r_{surf}$. In equation (\ref{temp}) we set
\begin{equation}
q_{rad}(r_{0}) = \frac{GM\dot{M}_{p}}{4\pi r_{0}^{2} r_{surf}}.
\label{qrad0}
\end{equation}
At each altitude we calculate pebble settling speed and pebble SiO loss, along with the atmosphere temperature, pressure, density, and SiO concentration using equations (\ref{ahydro})-(\ref{peqvp2}). 

Once the silicate pebble reaches the protoplanet surface, we calculate mass balances for SiO using equations (\ref{PvB})-(\ref{MvB}). The residual to equation (\ref{AvB}), representing anomalous volatile accumulation in the atmosphere, is used to correct the atmosphere saturation factor, and the solution steps listed above are repeated until the mass balances converge to better than 0.5\%. This typically requires 5-10 such iterations when starting from scratch. The number of iterations required is markedly reduced when we evolve a growing protoplanet, because the solution from the previous mass step provides excellent starting values for the next mass step.  

Next, we use the converged SiO solution along with initial estimates of the saturation ratios for Na, K, and Zn to calculate their initial abundance profiles, mixing ratios, and mass balances. We then iterate the solution for each volatile, updating the saturation ratio of each until its mass balances are satisfied to within 0.5\%. Here again, several iterations are needed to converge the solution for each of these elements.  Note that we use the atmosphere physical structure previously obtained for SiO during these iterations, and we solve for the three volatiles in sequence, independently of each other.  We justify this sequential approach on the basis of the relatively low vapor abundances of Na, K, and Zn. These are never abundant enough in the atmosphere to appreciably alter physical properties such as atmosphere temperature, density, or viscosity. 

For pebble flight, we obtain good results using a second order predictor-corrector integration scheme with time steps around 10$^3$s, which requires approximately 2$\times 10^4$ such steps for a silicate pebble to settle from the starting radius to the surface of a 0.7$M_{E}$ protoplanet. The radial increments from these time steps are also used as the altitude grid for calculating the physical properties of the atmosphere. For calculations with protoplanet growth, we evolve from an initial mass (usually 0.2$M_{E}$) in increments of 2$\times 10^{-3} M_{E}$ at constant 
$\dot{M}_{p}$, repeating pebble flight and updating all variables and properties at each mass level. 

We terminate protoplanet growth to 0.8$M_{E}$, for several reasons. First, beyond this mass the lowermost atmosphere fills with SiO. Segregation of H$_2$ and He from SiO is likely in such a layer, an effect not included in our model.  Second, at larger masses, near-surface temperatures rise above the region of validity of the vapor pressure measurements that determine the $\gamma$-factors in Table 6.  Lastly, our previous studies of hafnium-tungsten systematics (Olson and Sharp, 2023) and metal-silicate partitioning (Olson et al., 2025) indicate that the likely pebble mass of proto-Earth was less than 0.8$M_E$.

To calculate the azimuthal wind, we map the atmosphere radial density profile onto a cylindrical ($s,\phi,z$) grid surrounding the protoplanet, compute the gas column density using Appendix equation (A11), and then calculate the potential vorticity, the $z$-vorticity, and the azimuthal $\phi$-velocity on that grid, using Appendix equations (A12) -(A14).  We note that strictly 2D potential vorticity yields very high velocity azimuthal winds. As shown in the Appendix, 3D effects are modeled by adding a term for streamline convergence to the 2D potential vorticity, which yields somewhat weaker azimuthal winds. 

\section{Results}
\subsection{Atmosphere structure profiles}
Figures \ref{Me3} and \ref{Me7} show inner atmosphere structure profiles as functions of altitude for protoplanet masses 0.3$M_E$ and 0.7$M_E$, respectively, calculated using the pebble and atmosphere properties in Tables 4-6, with SiO as the only vapor. Temperature, pressure, and density increase toward the protoplanet surface, with basal atmosphere temperatures of 1665K and 3455K, and basal atmosphere pressures of 3.4 and 32 bars in the two cases. Opacity also increases toward the surface, particularly at 0.7$M_E$, because at high temperatures the contribution from gas molecules supersedes the contribution from dust. In contrast, the atmosphere total heat flux increases with altitude, a consequence of the volumetric internal heating of the atmosphere by the settling pebbles. 
\begin{figure}[h!]
\begin{center}
\includegraphics[width=0.7\linewidth]{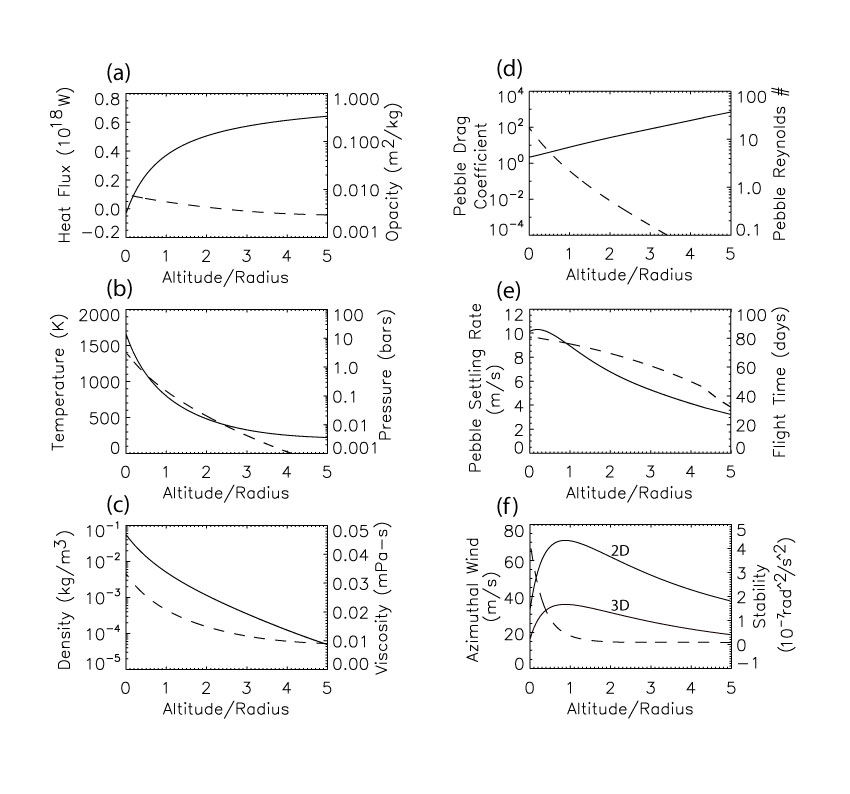}
\caption{Atmosphere structure profiles around a 0.3$M_E$ protoplanet growing by pebble accretion. Model variables and properties are given in Tables 3-6. In (a)-(f) the solid curves correspond to the left axes, the dashed curves to the right axes. Flight time begins at altitude 9$r_{surf}$. Azimuthal winds labeled 2D and 3D are described in the Appendix. Stability is $N^2$ from equation (\ref{Nsq}). }
\label{Me3}
\end{center}
\end{figure}

Figures \ref{Me3} and \ref{Me7} also show the pebble flight properties. As the pebbles enter the atmosphere they are initially subject to Epstein drag, which transforms to Stokes drag and then to quadratic drag as the pebbles approach the surface. During their descent, the pebble Reynolds number steadily increases, from ${\rm Re}<<$1 at high altitude to $\sim$100 near the surface. Similarly, the pebble settling rate increases from around 3 m/s at an altitude of 5$r_{surf}$ to 10 m/s in the near-surface at 0.3$M_E$, and to 6 m/s in the near-surface at 0.7$M_E$. The slight reduction in the near-surface settling rate with increasing protoplanet mass is primarily due to increased viscosity in the lowermost atmosphere at 0.7$M_E$, which more than compensates for the higher near-surface gravity. We note that relatively slow pebble settling rates translate into long flight times, and at large protoplanet masses, lengthy times over which the silicate pebbles are molten. For example, Figure \ref{Me3} shows that, at 0.3$M_E$, a pebble spends about 40 days to reach the surface from an altitude of  5$r_{surf}$. Similarly, Figure \ref{Me7} shows that at 0.7$M_E$, the silicate pebbles are fluid (above the rheological transition temperature $T_{rt}$) for more than 8 days. 
\begin{figure}[h!]
\begin{center}
\includegraphics[width=0.7\linewidth]{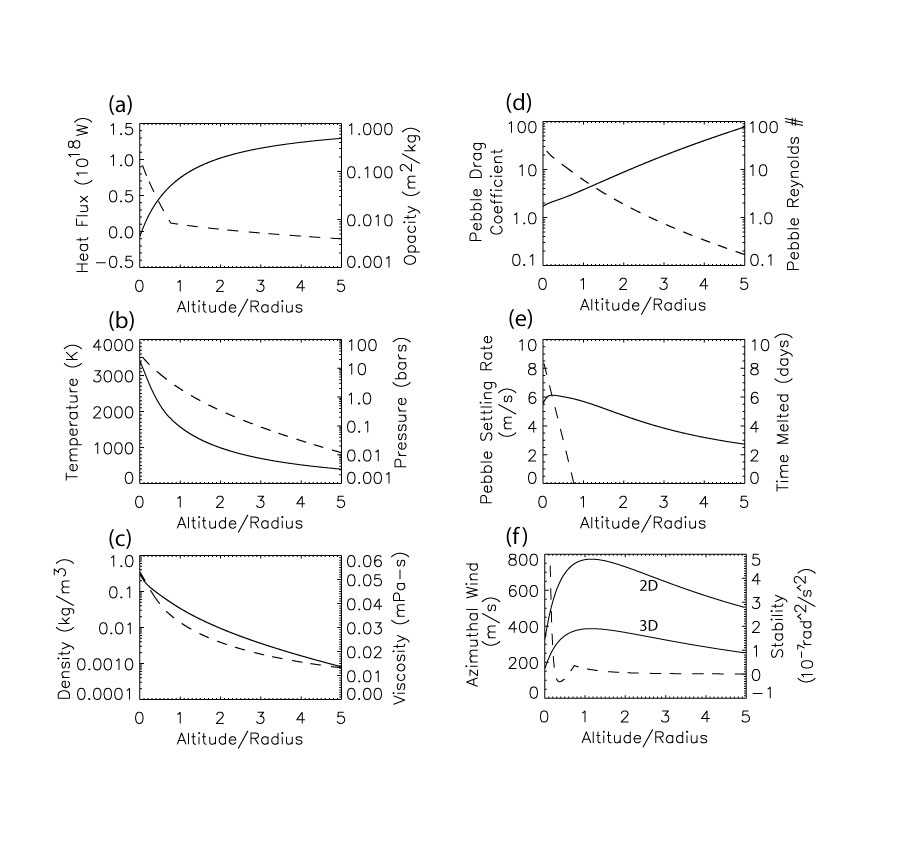}
\caption{Atmosphere structure profiles around a 0.7$M_E$ protoplanet growing by pebble accretion. Model variavles and properties are given in Tables 3-6. In (a)-(f) the solid curve corresponds to the left axes, the dashed curve to the right axes. Flight time begins at altitude 9$r_{surf}$. Azimuthal winds labeled 2D and 3D are described in the Appendix. Stability is $N^2$ from equation (\ref{Nsq}). }
\label{Me7}
\end{center}
\end{figure}

There are extremely large altitude variations in the atmosphere convective stability in Figures \ref{Me3}f and \ref{Me7}f. At high altitudes the atmosphere is essentially neutrally stable ($N^2\simeq$0) at 0.7$M_E$, and is very weakly stable at 0.3$M_E$.  Near the surface, strongly stable  conditions ($N^2>>$0) prevail in both cases, although for different reasons. The near-surface stable stratification at 0.7$M_E$ is due the concentration of SiO, in part from pebble vaporization but especially from vaporization at the protoplanet surface. In contrast, the near-surface stratification at 0.3$M_E$ is thermal in origin.
At 0.7$M_E$, a slightly unstably stratified ($N^2<$0) region develops just above the near-surface stable stratification, due to a change in the temperature gradient associated with the opacity transition from dust to gas-dominated.  Later we demonstrate that the structure of this unstable region is sensitive to the magnitude of the turbulent heat diffusivity $\kappa_{turb}$. 

Finally, we note that the azimuthal winds in Figures \ref{Me3}f and \ref{Me7}f far exceed the pebble settling rates, particularly in the 0.7$M_E$ case. Pebbles entering the atmosphere near the orbital plane of the protoplanet acquire high prograde velocities from the azimuthal wind, and execute multiple orbits during their decent. On reaching the surface, these fast-moving pebbles contribute prograde angular momentum to the protoplanet, which results in a bias for prograde rotation among protoplanets growing by pebble accretion (Johansen and Lacerda, 2010).
\begin{figure}[h!]
\begin{center}
\includegraphics[width=0.8\linewidth]{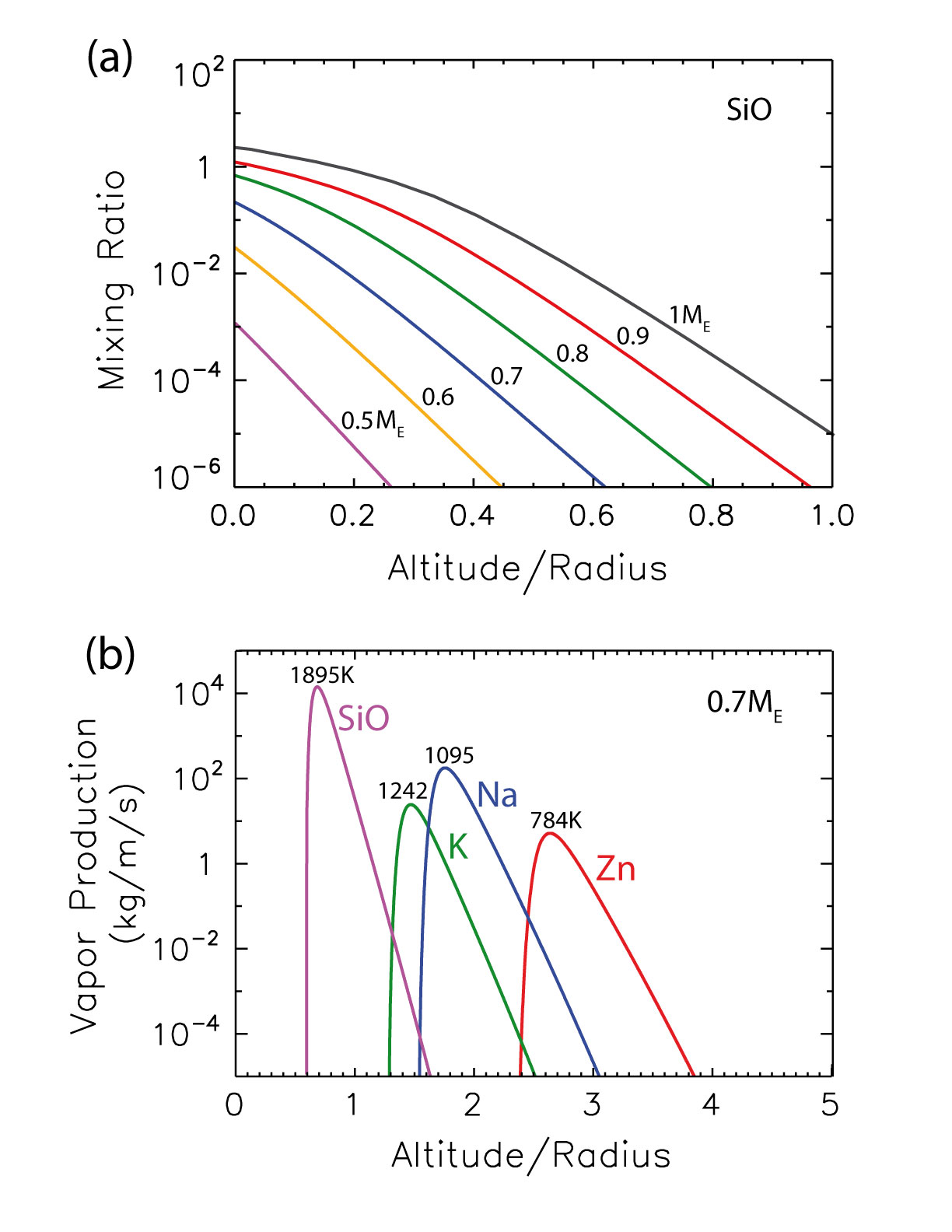}
\caption{(a): Atmosphere mixing ratios for SiO versus protoplanet mass, the silicate pebbles and the mantle having the silicate volatile abundances in Table 6. (b): Pebble vapor production versus altitude in a vapor-free nebular atmosphere above a 0.7$M_E$ protoplanet. Vaporization temperatures (peak vapor production temperatures) are indicated for each volatile.  }
\label{Vap}
\end{center}
\end{figure}

Figure \ref{Vap}a shows SiO mixing ratio profiles versus altitude for various protoplanet masses, the protoplanet mantles having the SiO abundance of the pebble model in Table 6. For protoplanet masses below 0.7$M_{E}$, the SiO mixing ratio increases exponentially on approach to the surface but remains much less than one at all altitudes. At increasingly larger masses, in contrast, the mixing ratio approaches and then exceeds one in the near-surface region. This behavior illustrates one of the previously-mentioned reasons for the protoplanet mass limitation we impose on our results: Above 0.8$M_{E}$, the near-surface SiO content of the atmosphere becomes high enough to drive segregation of the lighter H$_2$ and He components.

Figure \ref{Vap}b shows vapor production versus altitude from silicate pebbles with the initial volatile abundances given in Table 6, settling through a dry (i.e., volatile-free) atmosphere above a 0.7$M_{E}$ protoplanet. We note that a volatile-free atmosphere is not realistic under these conditions, because such an atmosphere would quickly approach volatile saturation. Nevertheless, this case is useful for illustrating where volatile loss from pebbles takes place. It reveals that the four volatiles vaporize in a discrete altitude bands, each altitude band defined by a narrow range of temperature.  Vapor production rapidly increases within each band, then decreases rapidly as vaporization becomes complete. The central temperature corresponding to each vaporization band is indicated in Figure \ref{Vap}b. Because of kinetic effects, these temperatures tend to be higher than the 50\% condensation temperatures in Table 1, although they are reasonably close for Na and K.  

\subsection{Volatiles with protoplanet growth}

\begin{figure}[h!]
\begin{center}
\includegraphics[width=0.6\linewidth]{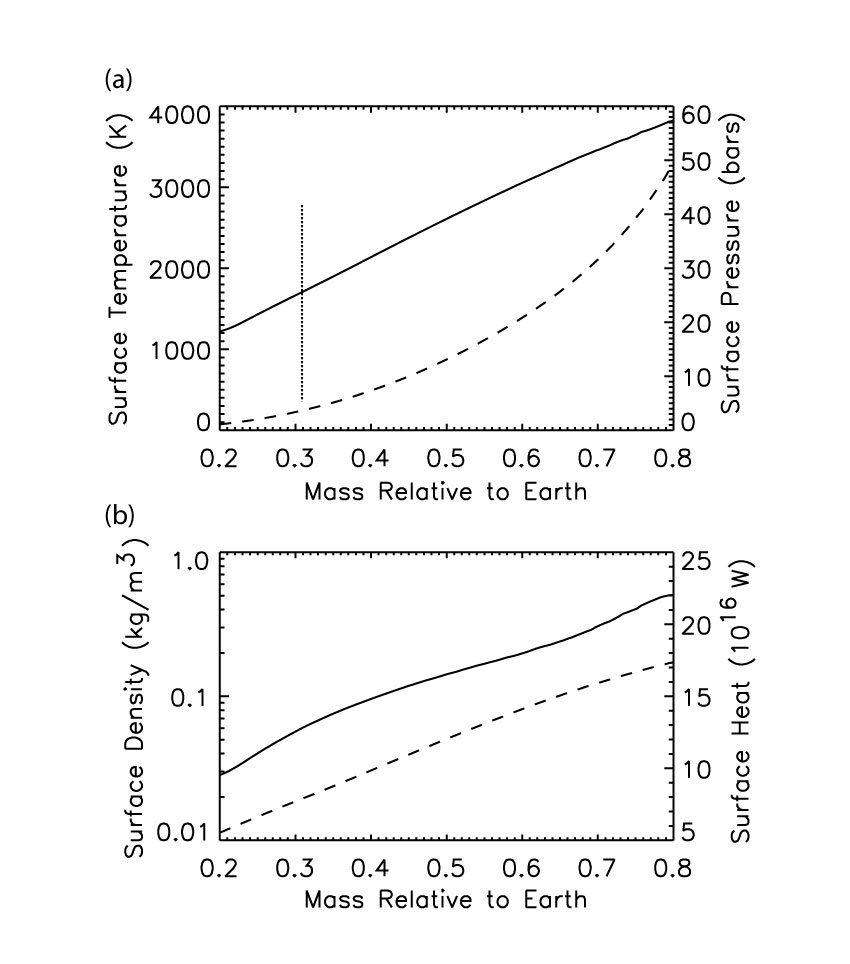}
\caption{Atmosphere temperature and pressure (a) and atmosphere density and sensible heat added at the protoplanet surface (b) versus protoplanet mass for the pebble accretion properties in Tables 4-6. The solid curves correspond to the left axes, the dashed curves to the right axes. The vertical dashed line denotes the mass beyond which the surface temperature exceeds the rheological transition temperature, permitting a surface magma ocean.}
\label{Surf1}
\end{center}
\end{figure}

Figure \ref{Surf1} shows the increases in surface temperature, pressure, density, and surface heat as a protoplanet grows steadily from 0.2$M_E$ to 0.8$M_E$. Here we have used the pebble and protoplanet properties in Tables 4-5, a 200 yr exhaust timescale, pebble volatiles from Table 6, with zero partitioning of volatiles into the core (i.e., $D_{cm}$=0). The vertical dotted line in Figure \ref{Surf1}a denotes the mass beyond which the surface temperature exceeds the rheological transition temperature and surface magma ocean conditions prevail. The inflection of the surface gas density curve in Figure \ref{Surf1}b around 0.65$M_E$ is a result of volatile addition. The surface heat in Figure \ref{Surf1}b is the rate at which sensible heat is added to the protoplanet by pebble deposition. It increases steadily with increasing protoplanet mass, in spite of the mass lost from the pebbles through partial vaporization. 
\begin{figure}[h!]
\begin{center}
\includegraphics[width=0.8\linewidth]{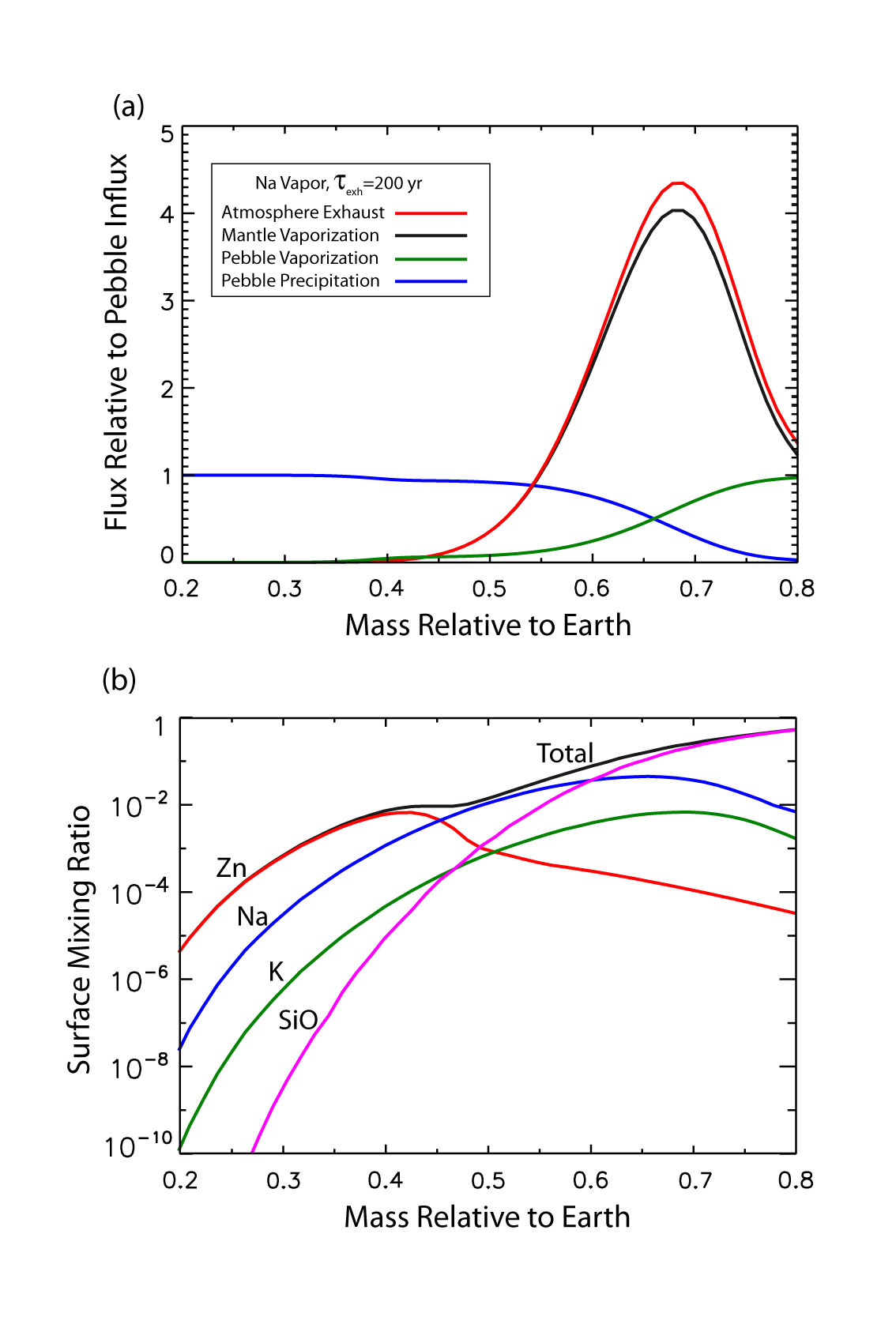}
\caption{(a): Fluxes of Na vapor normalized by pebble volatile influx versus protoplanet mass for pebble accretion properties in Tables 4-6.  (b): Atmosphere vapor mixing ratios at the surface versus protoplanet mass for the same case. }
\label{Surf2}
\end{center}
\end{figure}

Figure \ref{Surf2}a shows the contributions to the atmosphere mass balance for Na versus protoplanet mass for the same growth history as in Figure \ref{Surf1}. Vaporization and exhaust are negligible for Na below 0.3$M_E$, and only above 0.55$M_E$ do these become the dominant contributors. Pebble vaporization exceeds pebble precipitation above 0.65$M_E$. However, the contribution from mantle vaporization generally exceeds that from pebble vaporization, consistent with our assumption that the mantle, more than the pebbles, regulates the atmosphere volatile content. Figure \ref{Surf2}b shows the resulting mixing ratios at the surface for all four volatiles. For protoplanets smaller than 0.4$M_E$, Zn is the most abundant of the four. It is replaced by Na, which in turn is replaced by SiO starting around 0.6$M_E$. Note that the surface mixing ratios of Zn, K, and Na reach maxima at a particular mass, then decrease as the growing protoplanet becomes increasing depleted in these elements. We speculate that SiO behaves this same way, but in protoplanets larger than 0.8$M_{E}$. 

Figure \ref{EV2} shows the changes in volatile abundance versus mass for protoplanets growing by accretion of pebbles under the same conditions as Figures \ref{Surf1} and \ref{Surf2}, without volatile partitioning into the core (i.e., $D_{cm}$=0 in equation \ref{MvB}). The protoplanet volatile abundances in Figure \ref{EV2} are normalized by the pebble volatile abundances from Table 6 such that a value of one corresponds to zero volatile loss.  Note that all the abundance ratios in Figure \ref{EV2}a asymptote to one at small protoplanet masses, because vaporization and exhaust are negligible in this limit.  Beyond 0.3$M_E$, however, the abundance ratios decrease with increasing protoplanet mass, in order of their volatility. Zn is essentially depleted by 0.5$M_E$, Na and K are strongly depleted beyond 0.7$M_E$, while Si is only weakly depleted at 0.8$M_E$. 
\begin{figure}[h!]
\begin{center}
\includegraphics[width=0.6\linewidth]{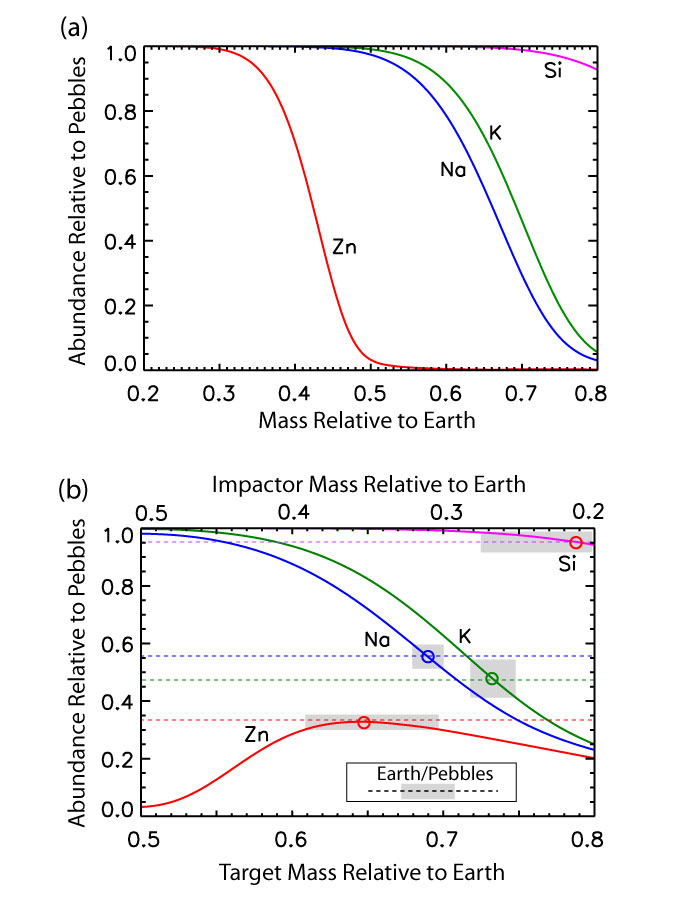}
\caption{ (a): Protoplanet volatile abundances normalized by pebble abundances versus protoplanet mass for the pebble accretion properties in Tables 4-6, without metal-silicate partitioning. (b): Normalized protoplanet volatile abundances for various 1$M_E$ target + impactor combinations from (a). Dashed lines and shadings are bulk Earth/pebble abundance ratios from Table 1 and bulk Earth uncertainties from Wang et al. (2018). Circles mark points of intersection. }
\label{EV2}
\end{center}
\end{figure}

Extrapolating the trends in a to larger mass, a 1$M_{E}$ protoplanet built this way would be substantially depleted in Si and practically devoid of K, Na, and Zn. 
However, there are several caveats to such an extrapolation. One is that some amount of each of these elements would have been sequestered in the protoplanet core, in which case the abundance ratios would not asymptote to zero at large protoplanet mass. We consider this effect later in this subsection. Another important caveat is the volatile contribution from large impacts. According to Figure \ref{EV2}a, protoplanets or planetary embryos smaller than approximately 0.3$M_E$ experience negligible loss of moderately volatile elements from pebble accretion. Therefore, impactors of this size or smaller are potential volatile sources for larger, depleted target protoplanets, particularly if the impactors were built from volatile-rich pebbles. 

This effect is illustrated in Figure \ref{EV2}b, which shows Si, K, Na, and Zn abundance ratios obtained by merging a target and an impactor from Figure \ref{EV2}a so as to produce a 1$M_E$ protoplanet. Here, the term merging signifies zero mass loss, including zero volatile loss, and we have assumed the impactor was built from the same pebble population as the target. This is reasonable as a starting assumption, because isotopic evidence indicates the proto-Earth and Moon-forming impactor Theia originated in the same part of the disk (Hopp et al., 2025). The dashed lines in Figure \ref{EV2}b are bulk Earth abundances normalized by the pebble abundances from Table 1, and the shadings indicate bulk Earth uncertainties according to Wang et al. (2018). Intersections of calculated and bulk Earth abundances are marked by circles. The agreement between calculated and Earth abundances for Zn, Na, and K is best for a target in the range 0.65-0.75$M_{E}$ and an impactor between 0.25 and 0.35$M_{E}$. The best agreement for Si requires a somewhat larger target and a somewhat smaller impactor, although the uncertainty for Si is great enough to so that its range partially overlaps with the ranges spanned by K. 
\begin{figure}[h!]
\begin{center}
\includegraphics[width=0.6\linewidth]{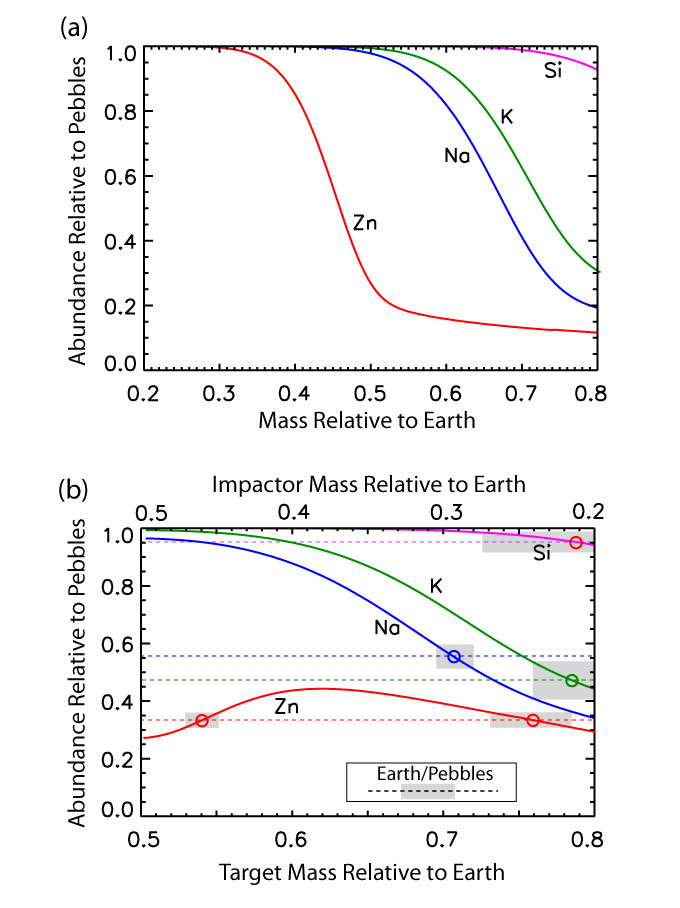}
\caption{ (a): Protoplanet volatile abundances normalized by pebble  abundances versus protoplanet mass for the pebble accretion properties in Tables 4-6, with metal-silicate partitioning included. (b): Normalized protoplanet volatile abundances for various 1$M_E$ target + impactor combinations from (a). Dashed lines and shadings are bulk Earth/pebble abundance ratios their uncertainties. Circles mark points of intersection. }
\label{EVD2}
\end{center}
\end{figure}

Figure \ref{EVD2} shows the effects of metal-silicate partitioning on volatile abundances from the same model as in  Figure \ref{EV2} but with core-mantle partitioning included via equation (\ref{MvB}) and  the $D_{cm}$-values from Table 6. Metal-silicate partitioning combined with volatile sequestration in the core causes the abundance ratios in Figure \ref{EVD2}a to decrease more slowly with mass, compared to Figure \ref{EV2}a. This slower decrease pushes the region of best agreement between calculated and Earth abundances toward somewhat larger targets and somewhat smaller impactors, as shown in Figure \ref{EVD2}b. 
Nevertheless, the regions of best agreement in Figures \ref{EV2} and \ref{EVD2} partially overlap, an indication that the exact amount of core-mantle partitioning is not crucial for estimating volatile loss. 
\begin{figure}[h!]
\begin{center}
\includegraphics[width=0.6\linewidth]{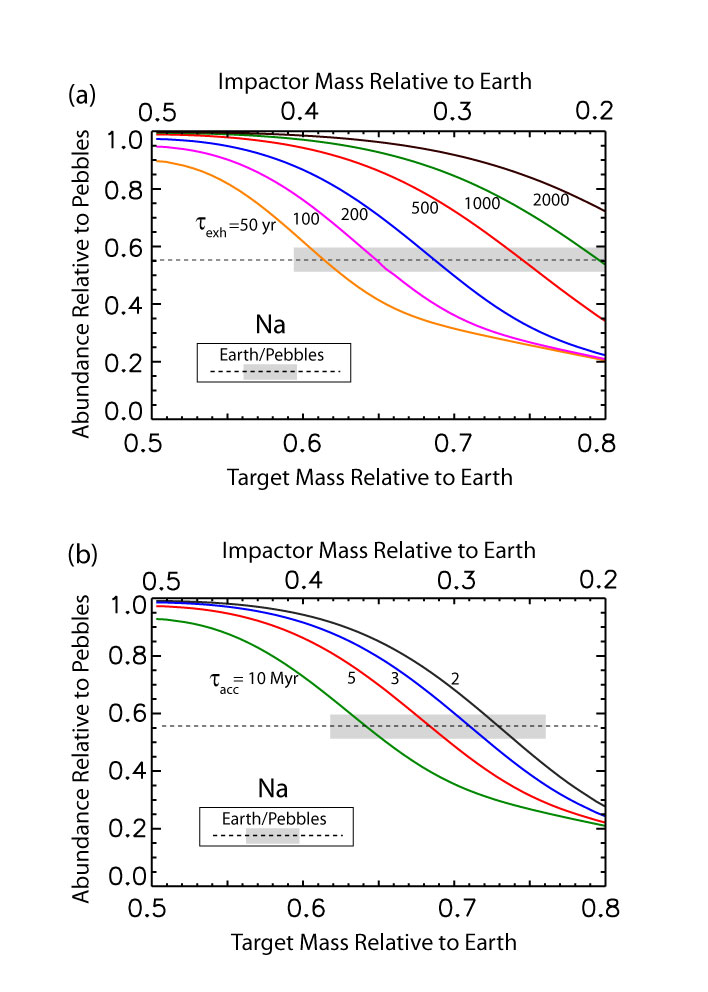}
\caption{Sensitivity of volatile depletion to atmosphere exhaust and pebble accretion timescales. (a): Sodium abundance ratios (protoplanet/pebbles) for various 1$M_E$ target + impactor combinations and various
atmosphere exhaust times $\tau_{exh}$. (b): Sodium abundance ratios for various accretion timescales $\tau_{acc}$. In both (a) and (b), the other model properties are the same as in Figure \ref{EV2}. Dashed line and shading are the bulk Earth/pebble sodium abundance ratio and its uncertainty.}
\label{ex}
\end{center}
\end{figure}

Figure \ref{ex} shows the sensitivity of the Na abundance ratio to variations in the atmosphere exhaust and the pebble accretion timescales, the other model properties being the same as in Figure \ref{EV2}. Figure \ref{ex}a shows how the Na abundance ratio changes with changes in the atmosphere exhaust timescale for different 1$M_{E}$ target + impactor combinations at a fixed pebble accretion timescale. Figure \ref{ex}b shows how the Na abundance ratio changes with the pebble accretion timescale at a fixed exhaust timescale. Agreement with Earth's Na abundance ratio is possible over a wide range of exhaust timescales $\tau_{exh}$, although for $\tau_{exh}>$1000 yr, the target protoplanet must be large ($\geq$0.8$M_E$) and the impactor(s) relatively small ($\leq$0.2$M_E$). Similarly, agreement with Earth is possible for a wide range of pebble accretion timescales $\tau_{acc}$. However, comparison of Figure \ref{ex}a and Figure \ref{ex}b indicates there is a tradeoff between the pebble accretion and the atmosphere exhaust timescales. If $\tau_{acc}$ is long, $\tau_{exh}$ must be short in order to match Earth's Na abundance ratio, and conversely, if $\tau_{acc}$ is short, $\tau_{exh}$ must be long. 
\begin{figure}[h!]
\begin{center}
\includegraphics[width=0.6\linewidth]{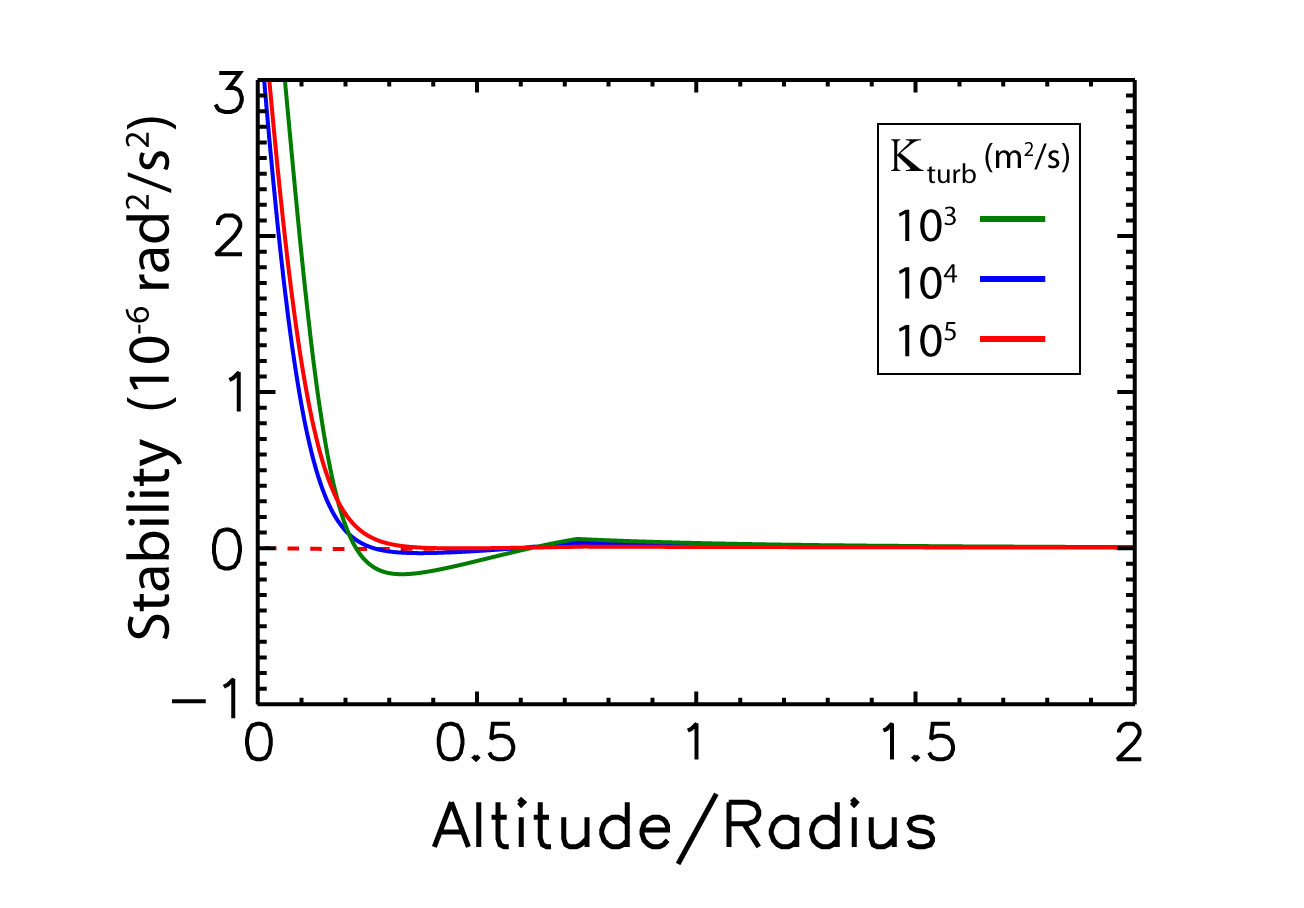}
\caption{Sensitivity of the atmosphere  stability parameter $N^2$ to various values of the turbulent diffusivity $\kappa_{turb}$, for a 0.7$M_E$ protoplanet growing by pebble accretion with properties in Tables 4-6. Zero marks neutral stability, with stable and unstable conditions above and below, respectively. The dashed line shows the stratification with $\kappa_{turb}$=$10^{5}$ m$^2$/s  with the effects of SiO vapor removed. }
\label{Kturb}
\end{center}
\end{figure}

Lastly, Figure \ref{Kturb} shows the atmosphere stability profiles $N^2$ that correspond to different values of the turbulent mixing parameter $\kappa_{turb}$ at 0.7$M_{E}$. The primary effect of increasing $\kappa_{turb}$ is to smooth the atmosphere temperature profile, and in doing so, smooth its stability profile. The smoothing effect is most evident in the altitude region between 0.2$r_{surf}$ and 0.6$r_{surf}$. This region is slightly unstable with $\kappa_{turb}=10^3$ m/s$^2$, essentially neutral with $\kappa_{turb}=10^4$ m/s$^2$ (our nominal value), and very slightly stable with $\kappa_{turb}=10^5$ m/s$^2$. The other parts of the stability profiles -- the near-surface stable stratification and the neutral stratification at higher altitude -- are hardly affected by variations in $\kappa_{turb}$. The high-altitude neutral stratification is consistent with convective turbulence in that region. The strong stable stratification in the near-surface region is a consequence of the increase in gas density due to abundant SiO vapor. Thermal effects alone would produce neutral stratification in the near-surface region. 
This is illustrated by the dashed line, which shows $N^2$ due to temperature variations, that is, without the effects of SiO vapor, for $\kappa_{turb}=10^5$ m/s$^2$. 

\section{Comparison with previous work}
A recent comprehensive investigation by Wang et al. (2025) of accretion-driven depletion of moderately volatile elements provides multiple points of comparison with our results. They use a 1D atmosphere structure model that incorporates the results of hydrodynamical simulations of atmosphere exhaust, similar to our general approach. They explain the differences in moderately volatile abundances on Earth and Mars in terms of a hybrid accretion history that combines volatile loss via atmosphere thermal processing under pebble accretion with volatile addition from one or more less-depleted giant impactors, plus contributions volatile-depleted planetesimals. Significantly, they find that the volatile contributions from depleted planetesimals are of secondary importance for matching Earth, but are critical for matching Mars. 

Although our general approach is broadly similar to theirs, the Wang et al. (2025) model details differ from ours in multiple ways.  In particular, their model uses CI chondrite volatile abundances for mantle-forming pebbles, enforces vaporization at the sublimation temperature for each volatile, includes a hot SiO vapor basal atmosphere layer, core sequestration based on equilibrium metal-silicate partitioning, and perhaps most importantly, atmosphere exhaust times of a few years for volatile elements. 

Yet in spite of these model differences, there is overall agreement between our results and theirs on certain key issues. In qualitative terms, both studies find that 1$M_{E}$ pebble-built planets are more depleted than Earth, whereas pebble-built planets much smaller than 1$M_{E}$ are less depleted than Earth. In quantitative terms, our best-fitting proto-Earth + impactor combinations in  Figures \ref{EV2} and \ref{EVD2} have mass combinations of (0.71 + 0.29)$M_{E}$ and (0.76 + 0.24)$M_{E}$, respectively, in good agreement with a best fit of (0.71 + 0.29)$M_{E}$ found by Wang et al. (2025) for proto-Earth plus a single giant impactor. This agreement is partly due to the fact that some of the differences between the two modeling approaches tend to compensate for each other, such as the tradeoff between pebble volatile abundance and atmosphere exhaust timescale, for example. But more fundamentally, the agreement is a consequence of the strong control that protoplanet mass has on volatile depletion during pebble accretion.

\section{Discussion}
Our results demonstrate that weakly and moderately volatile elements Si, Na, K, and Zn are thermally processed in the nebular atmosphere above a terrestrial protoplanet growing by pebble accretion. Significant vaporization of Na, K, and Zn from falling pebbles and also from the protoplanet surface begins above 0.3 Earth masses in our model. This combination of pebble and surface vaporization, coupled with atmosphere exhaust, indicates that large protoplanets accreting this way become depleted in moderately volatile elements compared to the pebbles that built them. Figure \ref{Comp} compares the depletion patterns of Earth and our best-fitting, most Earth-like models. 
\begin{figure}[h!]
\begin{center}
\includegraphics[width=0.6\linewidth]{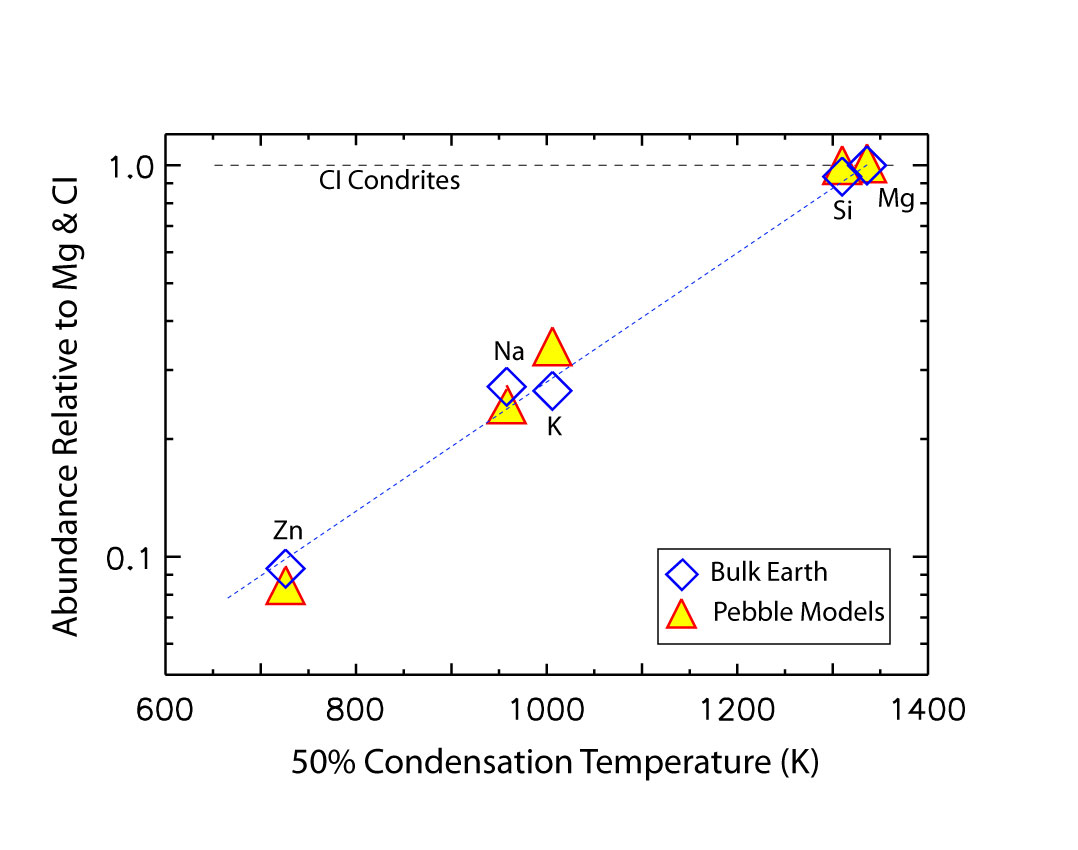}
\caption {Comparison of normalized element abundances versus condensation temperature for the bulk Earth (diamonds) and our best-fitting pebble accretion models (triangles) from
 Figures \ref{EV2} and \ref{EVD2} . These two best-fitting models are nearly identical on this plot. }
\label{Comp}
\end{center}
\end{figure}

The amount of moderately volatile depletion depends on multiple factors. In this study we have investigated three of these: the atmosphere exhaust and pebble accretion timescales, and the contributions from later impacts. We find that atmosphere exhaust times shorter than $10^3$ years coupled with pebble accretion timescales of 2 Myr or longer yield substantial depletion of moderately volatile elements on subEarth-size protoplanets, whereas atmosphere exhaust times greater than $10^3$ years result in far less depletion. In contrast, later mass addition by impacts tends to increase volatile abundances in cases where the impactors are too small to have been depleted by atmosphere thermal processing prior to collision with the target protoplanet.  In particular, using the pebble composition from Table 1, which matches Earth's major element abundances within uncertainties, we find that a partially depleted proto-Earth with mass near 0.7$M_{E}$ combined with one or more undepleted impactors with total mass near 0.3$M_{E}$ yields a depletion trend in broad agreement with the Earth for Si, Na, K, and Zn.

Volatile abundances in the input pebbles also affect our results.  Had we used the Mars data in Table 1 for our input pebble abundances, atmosphere thermal processing would yield more volatile depletion than the best-fitting target + impactor combinations in Figures \ref{EV2} and \ref{EVD2}.  Conversely, had we used the volatile-rich CI data in Table 1 for input pebble abundances, atmosphere thermal processing would yield less volatile depletion for these same target + impactor combinations. In short, for matching Earth's depletion trend, there is a tradeoff between the mass of the target protoplanet and the pebble volatile abundances. Volatile-rich pebbles require more atmosphere thermal processing to match Earth, which requires a larger target protoplanet, whereas volatile-poor pebbles require less thermal processing, which allows for a smaller target protoplanet.  

Another key assumption we make is that volatiles remain well-mixed in both the silicate pebbles and in the protoplanet mantle, allowing us to use their volume average concentrations when calculating equilibrium vapor pressures. This assumption is justified for pebbles at temperatures above their rheological transition temperature, because of the strong mixing effects from the circulation induced in the silicate liquid by atmospheric drag. It is more questionable in a protoplanet, however, because its internal circulation may not be vigorous enough to homogenize volatiles in its mantle. Without such homogenization, volatile concentrations at the surface might be determined by their concentrations in precipitating pebbles, for example, rather than by mantle average concentrations. 

Lastly, the absence of water vapor is perhaps the biggest shortcoming of our model. Water can be incorporated into protoplanets that migrate inward from beyond the snow line (Johansen et al., 2023a), or added by smaller impactors originating from beyond across the snow line (Albar\`{e}de, 2009; Raymond and Izidoro, 2017). Alternatively, water can be generated within the atmosphere itself, by hydrogen reduction of pebbles containing iron oxide (Garai et al., 2025b). Water can also be generated in magma oceans by this same reaction (Ikoma and Genda, 2006; Sharp, 2017; Olson and Sharp, 2019). Although a full examination of water in the context of pebble accretion is beyond the scope of this study, it has the potential to affect the results presented here.

The above complications and shortcomings notwithstanding, our results indicate that systematic depletion of moderately volatile elements is an expected outcome of pebble accretion in large terrestrial protoplanets, Earth-like planets in particular. And because volatile loss under pebble accretion depends so strongly on protoplanet mass, it is likely that the effects of this type of processing will be different in each planet. Accordingly, the details of pebble accretion-driven volatile loss need to be considered when comparing Earth to other terrestrial-type objects, including meteorite parent bodies. 

\section*{Acknowledgements} This research was partially supported by grant EAR1953992 from the National Science Foundation. We thank Anders Johansen for sharing a preprint in advance of publication. 

\section*{Data Statement} This paper is fully self-contained. All of the data needed to reproduce our results are included in the tables. No AI tools were used at any stage of this research. 

\section*{Author Contributions} PLO:  Investigation, Code Development and Validation, Writing.
ZDS: Project Support, Investigation; SG: Data Development and Curation, Proofreading.

\newpage
\clearpage

\section*{Appendix}

\subsection*{Atmosphere viscosity}
Let $\eta_{i}$ denote the viscosity of the $i^{th}$ gas, $X_{i}$ its mole fraction, and ${\rm m}_{i}$ its molecular weight. 
According to the Herning-Zipperer formulation (Monnery et al., 2016), the viscosity of the gas mixture is given by 
\begin{equation}
\eta =  \sum_{i} (X_{i}\eta_{i}/\sum_{j} X_{j}\Phi_{ij}),
\tag{A1}\label{eta}
\end{equation}
in which 
\begin{equation}
\Phi_{ij} = (\rm m_{j}/\rm m_{i})^{1/2}.
\tag{A2}\label{phi}
\end{equation} 
We use the following individual gas viscosity versus temperature profiles, all in Pa-s and degrees Kelvin. For hydrogen, from Stiel and Thodos (1963)
\begin{equation}
\eta_{H2} = 2.08\times10^{-6} (T/33.3)^{0.65};
\tag{A3}\label{etaH2}
\end{equation}
for helium, from Tournier and El-Genk (2008)
\begin{equation}
\eta_{He} = 3.063\times10^{-7} (T+21.3)^{0.7243};
\tag{A4}\label{etaHe}
\end{equation}
and for SiO, an extrapolation from Oh et al. (2018)
\begin{equation}
\eta_{SiO} = 5.82\times10^{-5} (T/1893)^{0.64}. 
\tag{A5}\label{etaHe}
\end{equation}

\subsection*{Outer atmosphere}
The outer nearly-isothermal atmosphere in our model begins at the Hill radius
\begin{equation}
r_{Hill}=(\frac{M}{3M_{S}})^{1/3}r_{\Omega},
\tag{A6}\label{rhill}
\end{equation}
where $M_{S}$ and $r_{\Omega}$ are Solar mass and protoplanet orbital radius, and ends at the top of the inner atmosphere at radius $r_{0}=10r_{surf}$ where the temperature is $T_{0}$.  
We specify that the temperature increases linearly with radius
through the outer layer according to $T=T_{n} +  (T_{0}-T_{n})(r_{Hill}-r)/(r_{Hill}-r_{0})$, where $T_{n}$ is the nebula temperature, and that the pressure varies according to 
\begin{equation}
\frac{dP}{dr}=-\frac{gP}{R'T},
\tag{A7}\label{Pout}
\end{equation}
starting from a nebula pressure $P_{n}$. Values for $T_{n}$, $P_{n}$, and $T_{0}$ are given in Table 5.

\subsection*{Azimuthal wind}
The strong azimuthal wind surrounding the protoplanet is a by-product of the compression of nebular gas entering the atmosphere, which amplifies the far-field vorticity of the disk gas. 
A simplified model of the azimuthal wind can be constructed by considering the steady flow around a protoplanet embedded in a gas-filled Keplerian disk, the protoplanet orbiting with angular velocity $\Omega$, as depicted in Figure \ref{Sketch1}.  In a co-revolving ($s,\phi, z$) cylindrical coordinate system centered on the protoplanet with $\phi$ increasing in the counterclockwise direction in Figure \ref{Sketch1}, the distribution of vorticity normal to the disk plane (that is, in the $z$-direction) for a barotropic, inviscid, and compressible gas is governed by Ertel's theorem (M\"{u}ller, 1995) with the form
\begin{equation} 
\frac{D}{Dt} \Gamma = 0,
\tag{A8}\label{A8}
\end{equation}
in which
\begin{equation} 
\Gamma =   (\frac{\omega + 2\Omega}{\Sigma})\frac{d\lambda}{dz}
\tag{A9}\label{A9}
\end{equation}
is the potential vorticity and $D/Dt$ denotes material derivative. Here, $\omega$ is the $z$-component of the vorticity of the gas, 
\begin{equation} 
\Sigma(s) =  \int_{-\infty}^{+\infty} \rho dz
\tag{A10}\label{A10}
\end{equation}
is the gas column density in the plane of the disk, and $\lambda$ is a scalar function that satisfies  
\begin{equation} 
\frac{D}{Dt}\lambda =0. 
\tag{A11}\label{A11}
\end{equation}

When  $\lambda=z$, the potential vorticity reduces to $\Gamma = (\omega + 2\Omega)/\Sigma$, the standard 
representation of potential vorticity for 2D disk flows around embedded protoplanets  (Korycansky and Papaloizou, 1996; Ormel, 2013). 
We can add 3D effects by including an axial displacement of the streamlines $\zeta$, so that $\lambda = z-\zeta$. We then
evaluate (A8) on a streamline, starting in the far-field (denoted by subscript $n$) where $\omega_{n}=-3\Omega/2$ and ending in the atmosphere at distance $s$. The result is $\Gamma(s)=\Gamma_{n}$, or, in
terms of (A9) and (A10),
\begin{equation}
\frac{(\omega + 2\Omega)(1-\zeta')}{\Sigma} = \frac{\Omega(1-\zeta'_{n})}{2 \Sigma_{n}}, 
\tag{A12}\label{A12}
\end{equation}
in which $\zeta'=d\zeta/dz$, $ \zeta'_{n}$ is its far-field value, and $\Sigma_{n}$ is the far-field column density. 
Setting $ \zeta'_{n}=0$ and rearranging (A12), the atmosphere vorticity becomes
\begin{equation}
\omega(s) = (\frac{\Omega}{1-\zeta'})(\frac{\Sigma(s)}{2\Sigma_{n}}) - 2\Omega.
\tag{A13}\label{A13}
\end{equation}
Using Kelvin's circulation theorem, the azimuthal wind surrounding the protoplanet is given in terms of (A13) as
\begin{equation}
v_{\phi}(s) = \frac{1}{s}\int_{\hat{s}=0}^{s} \hat{s}\omega(\hat{s}) d\hat{s}.
\tag{A14}\label{A14}
\end{equation}

Note that the azimuthal wind $v_{\phi}$ is columnar (i.e., cylindrically symmetric, variable along $s$, and invariant along $z$) and therefore does not, in general, conform to solid body rotation of a nearly spherical protoplanet.  Nevertheless, equation (A14) gives the approximate azimuthal wind pattern experienced by pebbles circulating in or near the orbital plane of the protoplanet (the plane defined by $z=0$) as they settle through the atmosphere. 

To calculate the azimuthal wind, we map the radial atmosphere density profile between $r_{surf}$ and 10$r_{surf}$ onto a cylindrical ($s,z$)-grid, on which we compute the gas column density $\Sigma(s)$ using (A10). We then compute the vorticity $\omega(s)$ using (A13), and from that, the azimuthal wind $v_{\phi}(s)$ using (A14) with $v_{\phi}(0)=$0. Setting $\zeta'$=0 yields the 2D wind; setting $\zeta'<0$ adds the effect of gas convergence toward the disk mid-plane, and reduces the azimuthal wind relative to 2D cases. The azimuthal winds shown in Figures \ref{Me3} and \ref{Me7} use $\Sigma_{n}$= 1000 kg/m$^2$; those labeled 2D use $\zeta'$=0 and those labeled 3D use $\zeta'=-1$.

\newpage

\section*{References}
 
\begin{description}

\item Adolfsson, L.G., Gustafson, B.A.S., Murray, C.D., 1996. The martian atmosphere as a meteoroid detector. Icarus 119, 144-152.

\item Albar\`{e}de, F., 2009. Volatile accretion history of the terrestrial planets and dynamic implications.
Nature 461, 1227-1233.

\item Bethune, W., Rafikov, R.R., 2019a. Envelopes of embedded super-Earths - II. Three-dimensional simulations. Mon. Not. Roy. Astron. Soc. 488, 2265-2379. 

\item Bethune, W., Rafikov, R.R., 2019b. Envelopes of embedded super-Earths - I. Two-dimensional simulations. Mon. Not. Roy. Astron. Soc. 487, 2319-2334. 

\item Bell, K.R., Lin, D.N.C., 1994. Using FU Origins outbursts to constrain self-regulated protostellar disk models. Astrophys. J. 427, 987.

\item Bizzarro, M., Johansen, A., Dorn, C., 2025. The cosmochemistry of planetary systems. 
       Nature Rev. Chem. https://doi.org/10.1038/s41570-025-00711-9.
       
\item Braukm\"{u}ller, N., Wombacher, F., Funk, C., M\"{u}nker, C., 2019. Earth's volatile element depletion
     pattern inherited from a carbonaceous chondrite-like source. Nature Geosci. 12, 564-568.

\item Brouwers, M.G., Vazan, A., Ormel, C.W., 2018. How cores grow by pebble accretion 1. Direct core growth. Astron. Astrophys. 611, A65.  

\item Chambers, J., 2023. Making the Solar System. Astrophys. J. 944:127, 23pp. 

\item Colmenares, M. J., Lambrechts, M., van Kooten, E., Johansen, A., 2024. Thermal processing of primordial pebbles in evolving protoplanetary disks. Astron. Astrophys. 685, A114.

\item Dias, B., Turchi, A., Stern, E.C., Magin, T.E., 2020. A model for meteorite ablation including melting and vaporization.
Icarus 345, 113710. 

\item Garai, S., Olson, P., Sharp, Z., 2025a. Building Earth from pebbles made of chondritic components.  Geochim. Cosmochim. Acta 390, 86-104.
	
\item Garai, S., Sharp, Z.D., Olson, P.L., Gargano, A.M., 2025b. Chondritic component pebbles as sources of Earth's composition and water. 
https://doi.org/10.7185/gold2025.26067.
	
\item Grewal, D.S., Miyazaki, Y., Nice, N.X., 2024. Contribution of the Moon-forming impactor to the volatile inventory in the bulk silicate Earth. Planet. Sci. J. 5:181, 11pp. 	
	
\item Hill, K.A., Rodgers, L.A., Hawkes, R.L., 2005. High geocentric velocity meteor ablation. Astron. Astrophys. 444, 615-624. 

\item Hin, R.C., Coach, C.D., Carter, P.J., Nimmo, F. et al., 2017. Magnesium isotope evidence that accretion vapor loss shapes planetary compositions. Nature 549, 511-515.
	
\item Hopp, T., Dauphas, N., Boyet, M. et al., 2025. The Moon-forming impactor Theia originated from the inner Solar System. Science doi:10.1126/science.ado0623.

\item Ikoma, M., Genda, H., 2006. Constraints on the mass of a habitable planet with water of nebular origin. Astrophys. J. 648, 696.  

\item Johansen, A., Ronnet, T., Schiller, M., Deng, Z., Bizzarro, M., 2023a. Anatomy of rocky planets formed by rapid pebble accretion I. 
          How icy pebbles determine the core fraction and Fe contents.  Astron. Astrophys. 671, A74.

\item Johansen, A., Ronnet, T., Schiller, M., Deng, Z., Bizzarro, M., 2023b. Anatomy of rocky planets formed by rapid pebble accretion II. 
     Differentiation by accretion energy and thermal blanketing.  Astron. Astrophys. 671, A75.

\item Johansen,  A., Ronnet,  T., Bizzarro, M., Schiller, M., Lambrechts M., Nordlund, Å., Lammer H., 2021. A pebble accretion model for the formation of the  terrestrial planets in the Solar System. Science Advances 7, eabc0444.
          
\item Johansen, A., Nordlund, Å., 2020. Transport, destruction and growth of pebbles in the gas envelope of a protoplanet. Astrophys. J. 903, 102 (13pp).

\item Johansen, A., Lacerda, P., 2010. Prograde rotation of propoplanets by accretion of pebbles in a gaseous environment. Mon. Not. Roy. Astron. Soc. 404, 475-485.

\item Kitts, K., Lodders, K., 1998. Survey and evaluation of eucrite bulk compositions. Meteror. Planet. Sci. 33, A197-A213. 

\item Korycansky, D.G., Papaloizou, J.C.B., 1996. A method for calculations of nonlinear shear flow: application to formation of giant planets in the solar nebula. Astrophys. J. Suppl. Ser. 105, 181-190.

\item Kurokawa, H., Tanigawa, T., 2018. Suppression of atmospheric recycling of planets embedded in a protoplanetary disc by buoyancy barrier. Mon. Not. Roy. Astron. Soc. 479, 635-648.

\item Levison, H. F., Kretke, K. A., Walsh, K. J., Bottke W. F., 2015. Growing the terrestrial planets from the gradual accumulation of submeter-sized objects.  Proc. Natl. Acad. Sci. 112, 14180-14185.
        
\item Li, J., Bergin, E.A., Blake, G.A., Ciesla, F.J., Hirschmann, M.M., 2021. Earth's carbon deficit caused by early loss through irreversible sublimation. Science Advances 7:14, eabd3632.

\item Liou, K.-N., Ou, S.-C. S., 1983. Theory of equilibrium temperatures in radiative-turbulent atmospheres. J. Atmos. Sci. 40, 214-229.

\item Lock, S.J., Stewart, S.T., 2024. Atmospheric Loss in Giant Impacts Depends on Pre-impact Surface Conditions. Planet. Sci. J. 5:28, 35pp. 

\item Lodders, K., 2021. Relative atomic solar system abundances, mass fractions, and atomic masses of the elements and their isotopes, composition of the solar photosphere, and compositions of the major chondritic meteorite groups. Space Sci. Rev. 217:44. 
                      
\item Lodders, K., 2003. Solar system abundances and condensation temperatures of the elements. Astrophys. J. 591, 1220-1247. 

\item McDonough, W.F., 2025. Earth's composition, origin, evolution and energy budget. arXiv preprint arXiv:2505.02641.

\item Moldenhauer, T.W., Kuiper, R., Kley, W., Ormel, C.W., 2021. Steady state by recycling prevents premature collapse of protoplanetary atmospheres. Astron. Astrophys. 646, L11. 

\item Moldenhauer, T.W., Kuiper, R., Kley, W., Ormel, C.W., 2022. Recycling of the first atmospheres of embedded planets: Dependence on core mass and optical depth. Astron. Astrophys. 661, A142.

\item Monnery, W.D., Svrcek, W.Y., Mehrotra, A.L., 2016. Viscosity: A critical review of practical predictive and correlative methods. Can. J. Chem. Eng. 73, 3-40. 

\item Morbidelli, A., Kleine, T., Nimmo, F., 2025. Did the terrestrial planets of the solar system form by pebble accretion? Earth Planet. Sci. Lett. 650, 119120. 

\item Morbidelli A., Nesvorny D., 2012. Dynamics of pebbles in the vicinity of a growing planetary embryo: hydrodynamical simulations. Astron. Astrophys. 546:A18.

\item M\"{u}ller, P., 1995. Ertel's potential vorticity theorem in physical oceanography. Rev. Geophys. 33:1, 67-97.

\item N\"{a}slund, E., Thaning, L., On the settling velocity in a nonstationary atmosphere. Aerosol Sci. Tech. 14, 247-256.

\item Oh, J.S., Kim, H.S., Lee, J. et al., 2018. Kinetics of dissolution of SiO gas in liquid Fe-C alloys. ISIJ International 58:12, 2246-2252. 

\item Olson, P.L., Sharp, Z.D., Garai, S., 2025. Pebble accretion and siderophile element partitioning between Earth's mantle and core. Phys. Earth Planet. Inter. 358, 107295.

\item Olson, P.L., Sharp, Z.D., 2023. Hafnium-tungsten evolution with pebble accretion during Earth formation. Earth Planet. Sci. Lett. 622, 118418.

\item Olson P.L., Sharp Z.D., 2019. Nebular atmosphere to magma ocean: A model for volatile capture during Earth accretion. Phys. Earth Planet. Inter. 294, 106294.

\item Onyett, I.J, Schiller, M., Makhatadze, G.V. et al., 2023. Silicon isotope constraints on terrestrial planet accretion. 
               Nature https://doi.org/10.1038/s41586-023-06135-z.
               
\item Ormel, C.W., Shi, J.-M., Kuiper, R., 2015. Hydrodynamics of embedded planets' first atmospheres II. A rapid recycling of atmospheric gas. Mon. Not. Roy. Astron. Soc. 447, 3512-3525.

\item Ormel, C.W., Klahr, H.H., 2010. The effect of gas drag on the growth of protoplanets. Analytical expressions for the accretion 
of small bodies in laminar disks. Astron. Astrophys. 520, A43, 15pp.

\item Ormel, C.W., 2013. The steady-state flow pattern past gravitating bodies. Mon. Not. Roy. Astron. Soc. 428, 3526-3542.

\item Popovas, A., Nordlund, Å., Ramsey, J.P., Ormel, C.W., 2018. Pebble dynamics and accretion on to rocky planets - I. Adiabatic and convective models. Mon. Not. Roy. Astron. Soc. 479, 5136-5156. 

\item Raymond, S.N., Izidoro, A., 2017. Origin of water in the inner Solar System: planetesimals scattered inward during Jupiter and Saturn?s rapid gas accretion. Icarus 297, 134-148. https://doi.org/10.1016/j.icarus.2017.06.030.

 \item Schiller, M., Bizzarro, M., Siebert, J., 2020.  Iron isotope evidence for very rapid accretion and differentiation of the proto-Earth. Science Advances 6, eaay7604.

\item Sharp, Z.D., 2017. Nebular ingassing as a source of volatiles to the terrestrial planets. Chem. Geol. 448, 137-150.

\item Sharp, Z.D., Olson, P.L., 2022. Multi-element constraints on the sources of volatiles to Earth. Geochem. Cosmochim. Acta 333, 124-135. 

\item Sossi, P.A., Stotz, I.L., Jacobson, S.A., Morbidelli, A., O'Neill, H.S.C., 2022. Stochastic accretion of the Earth. Nature Astron. 6, 951-960. https://doi.org/10.1038/s41550- 022-01702-2. 
 
\item Steinmeyer, M.-L., Johansen, A., 2024. Vapor equilibrium models of accreting rocky planets demonstrate direct core growth by pebble accretion. Astron. Astrophys. 683, id.217. 

\item Steinmeyer, M.-L., Woitke, P., Johansen, A., 2023. Sublimation of refractory minerals in the gas envelopes of accreting rocky planets. Astron. Astrophys. 677, id.A181, 16pp.  

\item Stiel, L.I., Rhodos, G., 1963. Viscosity of hydrogen in the gaseous and liquid states for temperatures up to 5000$^o$ K. Ind. Eng. Chem. Fundamentals v:3. 223-237.

\item Tian, Z., Chen, H., Fegley, B. Jr., Lodders, K. et al., 2019. Potassium isotopic compositions of howardite-eucrite-diogenite meteorites. Geochim. Cosmochim, Acta 266, 611-632. 

\item Tournier, J.M.P., El-Genk, M.S., 2008. Properties of helium, nitrogen and He-N$_2$ binary gas mixtures. J. Thermophys. Heat Trans. 22:3, 442-456.

\item Wang, H.S., Johansen, A., Xu, Z., Steinmeyer, M.-L. et al., 2025. Hybrid accretion of rocky planets imprinted in volatile depletion. Nature Astron. (submitted). 

\item Wang, H.S., Lineweaver, C. H., Ireland, T. R., 2018. The elemental abundances (with uncertainties) of the most Earth-like planet. Icarus 299, 460-474.

\item Wang, Y., Ormel, C.W., Huang, P., 2023. Atmospheric recycling of volatiles by pebble-accretion. Mon. Not. Roy. Astron. Soc. 523, 6186-6207. 

\item Yoshizaki T., McDonough, W.F., 2020. The composition of Mars. Geochim. Cosmochim. Acta 273, 137-162.

\end{description}

\clearpage
\newpage

\end{document}